\documentclass[conference]{IEEEtran}
\IEEEoverridecommandlockouts
\usepackage{cite}
\usepackage{amsmath,amssymb,amsfonts}
\usepackage{graphicx}
\usepackage{textcomp}
\usepackage{xcolor}
\usepackage{algorithm}
\usepackage{algpseudocode}
\usepackage{subcaption}
\usepackage{tabularx}
\usepackage{comment}
\usepackage{caption}
\let\oldalgorithm\algorithm
\renewcommand{\algorithm}{
    \oldalgorithm
    \footnotesize
}
\def\BibTeX{{\rm B\kern-.05em{\sc i\kern-.025em b}\kern-.08em
    T\kern-.1667em\lower.7ex\hbox{E}\kern-.125emX}}
\begin{document}

% \title{sTiles: A Sparse-Dense Tile-Based Framework for Structured Matrices with GPU Acceleration}
\title{\emph{sTiles}: An Accelerated Computational Framework for Sparse Factorizations of Structured Matrices}
\author{}

%\begin{comment}
    \author{
\IEEEauthorblockN{Esmail Abdul Fattah}
\IEEEauthorblockA{\textit{CEMSE Division} \\
\textit{King Abdullah University of Science and Technology}\\
Thuwal, 23955, Makkah, Saudi Arabia \\
esmail.abdulfattah@kaust.edu.sa}
\and
\IEEEauthorblockN{Hatem Ltaief}
\IEEEauthorblockA{\textit{CEMSE Division} \\
\textit{King Abdullah University of Science and Technology}\\
Thuwal, 23955, Makkah, Saudi Arabia \\
hatem.ltaief@kaust.edu.sa}
\and
\IEEEauthorblockN{H{\aa}vard Rue}
\IEEEauthorblockA{\textit{CEMSE Division} \\
\textit{King Abdullah University of Science and Technology}\\
Thuwal, 23955, Makkah, Saudi Arabia \\
haavard.rue@kaust.edu.sa}
\and
\IEEEauthorblockN{David Keyes}
\IEEEauthorblockA{\textit{CEMSE Division} \\
\textit{King Abdullah University of Science and Technology}\\
Thuwal, 23955, Makkah, Saudi Arabia \\
david.keyes@kaust.edu.sa}
}
%\end{comment}

\maketitle

% \textcolor{red}{PLEASE FEEL FREE TO REWRITE THE TITLE}\\

\begin{abstract}
This paper introduces \emph{sTiles}, a GPU-accelerated software framework that computes factorization of sparse structured symmetric matrices. By leveraging tile algorithms to achieve fine-grained computations, \emph{sTiles} employs a structure-aware flow of task executions for the Cholesky factorization to tackle challenging arrowhead sparse matrices with variable bandwidths. These matrices, common in various scientific and engineering fields, necessitate an adaptive factorization strategy that operates only on nonzero tiles. Our approach relies first on a collection of permutation techniques to minimize fill-in during factorization. Then, \emph{sTiles} proceeds with a customized static scheduler to orchestrate computational tasks on shared-memory systems equipped with GPU hardware accelerators. \emph{sTiles} strikes the required balance between the tile size and the degree of parallelism. The former has a direct impact on the algorithmic intensity, although monitoring the additional floating-point operations and memory footprint becomes crucial for parallel performance. The latter may be limited due to the inherent arrowhead sparse matrix structure. To further expose parallelism, we must adopt a left-looking variant of the Cholesky factorization to break the sequential dependencies during the accumulation operations on the trailing submatrix and operate on them instead using tree reductions. Extensive evaluations demonstrate a boost in performance for the sparse Cholesky factorization on various arrowhead structured matrices against the state-of-the-art sparse libraries, with \emph{sTiles} achieving up to 8.41X/9.34X/5.07X/11.08X speedups compared to CHOLMOD/SymPACK/MUMPS/PARDISO, respectively. On GPUs, \emph{sTiles} further exploits the computational throughput by adjusting the two key ingredients aforementioned to attain a 5X speedup compared to a 32-core AMD EPYC CPU when executed on an NVIDIA A100 GPU. Our generic software framework imports well-established concepts from dense matrix computations but they all require customizations in their deployments on hybrid architectures to best handle factorizations of sparse matrices with arrowhead structures.
\end{abstract}

\section{Introduction}

Block arrowhead matrices are widely studied in various scientific and engineering disciplines due to their exploitable structure, in which nonzero elements are primarily located in the last block row, the last block column, and along the block diagonal. This configuration facilitates specific mathematical operations, minimizing fill, which can be particularly advantageous in solving large-scale problems. Their application across mathematics, physics, and engineering highlights the practical significance of these matrices and the potential to advance computational methods by addressing this pattern. Notable tasks required with block arrowhead matrices include computing spectral decomposition \cite{stor2015accurate, diele2004computing}, factorizing posterior matrices \cite{gaedke2024integrated}, solving symmetric arrowhead matrices \cite{beik2015iterative}, computing their determinants \cite{hetmaniokdeterminants}, computing their inverses \cite{holubowski2015fast, gaedke2024integrated}, and many others.

\begin{figure}%[htbp]
    \centering
    \begin{minipage}[b]{0.3\columnwidth}
        \centering
        \includegraphics[width=\columnwidth]{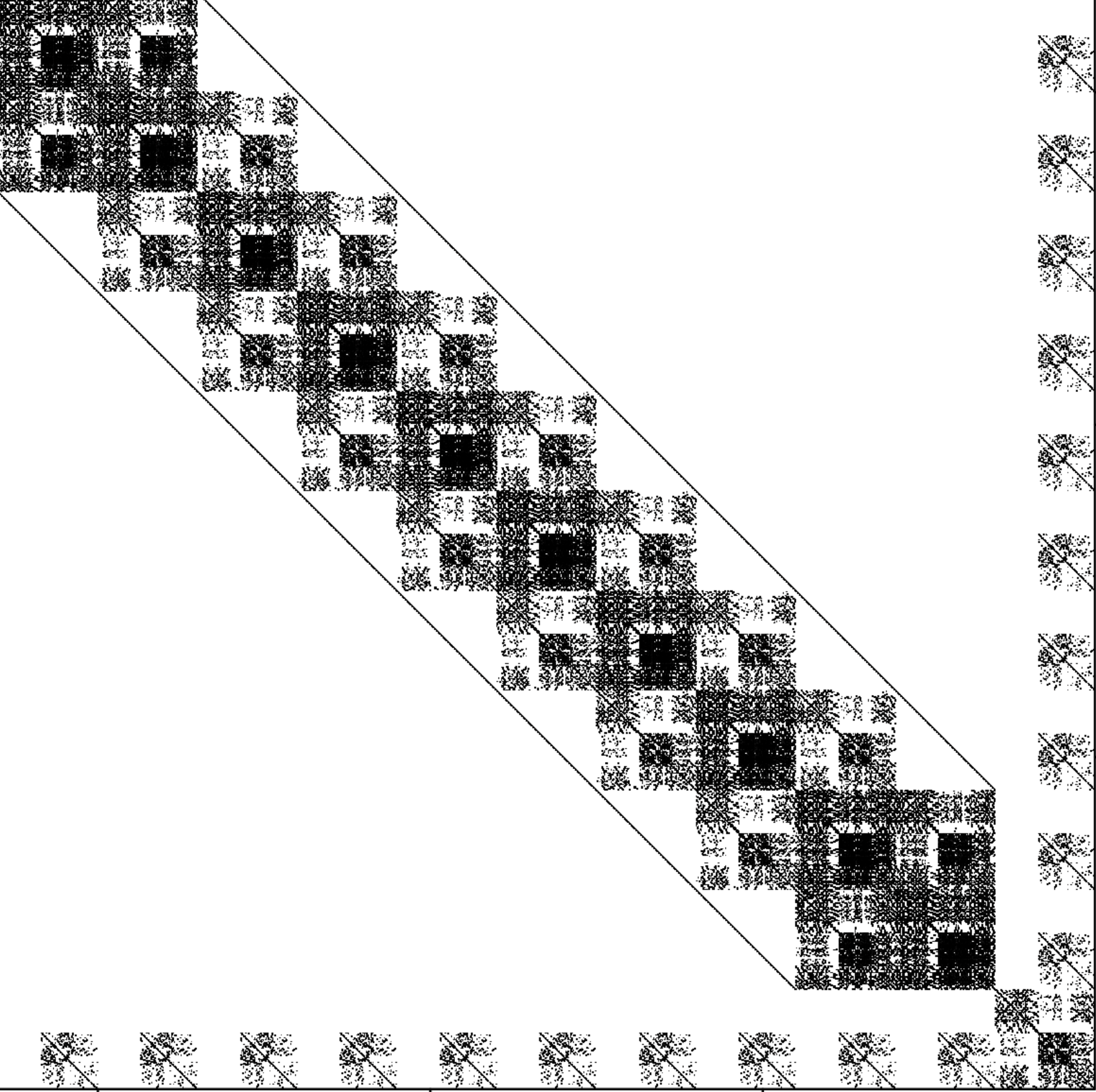}
    \end{minipage}
    \hspace{0.01\columnwidth} % Reduced spacing
    \begin{minipage}[b]{0.3\columnwidth}
        \centering
        \includegraphics[width=\columnwidth]{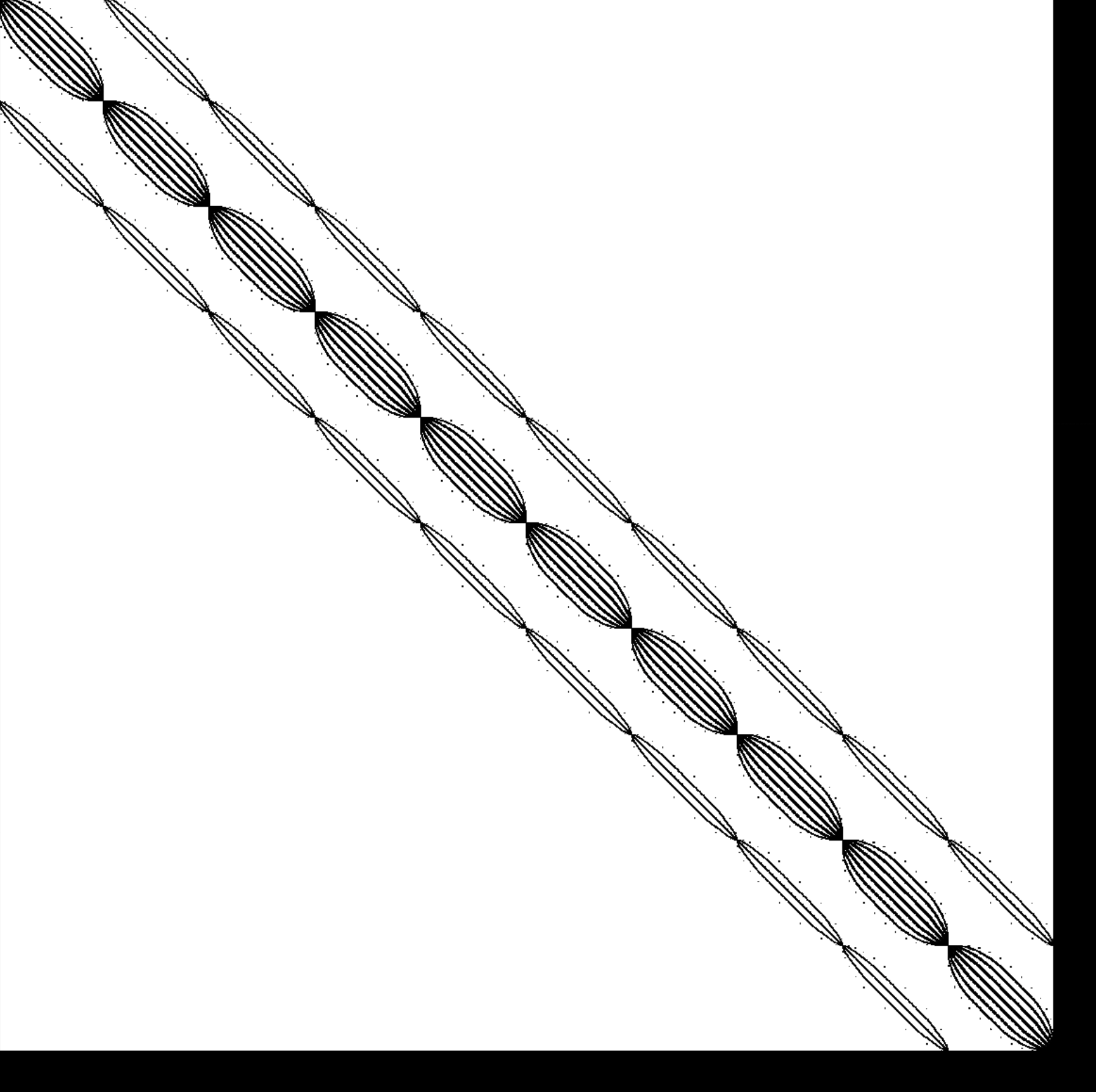}
    \end{minipage}
    \hspace{0.01\columnwidth} % Reduced spacing
    \begin{minipage}[b]{0.3\columnwidth}
        \centering
        \includegraphics[width=\columnwidth]{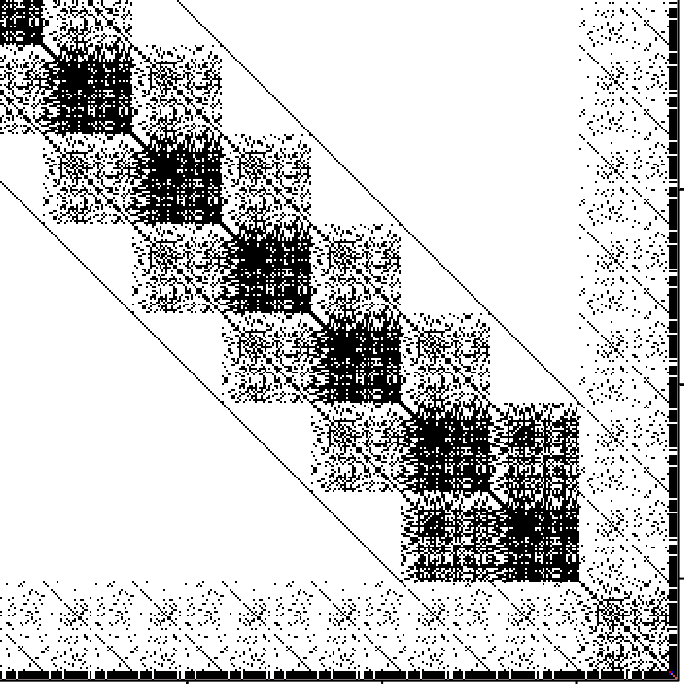}
    \end{minipage}
    \caption{Matrix patterns for different Bayesian inference applications.}
    \label{patternsAHM}
\end{figure}

These types of matrices frequently arise in Bayesian inference, particularly in spatial and spatiotemporal modeling, due to their role in efficiently managing the high-dimensional latent parameter spaces inherent in these models, as shown in Figure \ref{patternsAHM}. The methodology of Integrated Nested Laplace Approximations (INLA) \cite{van2023new}, combined with the Stochastic Partial Differential Equations (SPDE) approach \cite{lindgren2011explicit}, makes extensive use of these matrices for large-scale Bayesian spatial-temporal modeling. INLA has been widely applied in various fields such as biology, environmental science, ecology, public health, biostatistics, geoscience, epidemiology, and meteorology, demonstrating its versatility in handling complex models. During the INLA inference process, Cholesky decomposition for block arrowhead matrices is needed frequently, and the number of decompositions can easily reach hundreds as the dimension of the hyperparameters grows.

Most traditional linear algebra libraries do not fully exploit the unique structural properties of block arrowhead matrices during Cholesky factorization, which results in missed opportunities for performance gain. Sparse solvers, while efficient for general sparse matrices, are not optimized for the specific structure of block arrowhead matrices, leading to low arithmetic intensity kernels despite the presence of dense blocks or clusters. At the opposite end, dense solvers treat all elements as nonzero, failing to leverage the sparsity within block arrowhead matrices, which leads to higher algorithmic complexity and memory footprint.

We introduce \emph{sTiles}, an accelerated computational software framework that promotes a hybrid approach, one that combines the benefits of both sparse and dense solvers. \emph{sTiles} transforms the dense blocks within block arrowhead matrices into a collection of smaller dense tiles before eventually performing sparse Cholesky factorization. By focusing computational efforts on these dense tiles where the nonzero elements are concentrated and skipping the sparse regions, \emph{sTiles} can significantly enhance performance, resulting in faster and more scalable software solutions.

To address the computational demands of block arrowhead
structured matrices, \emph{sTiles} deploys a customized version of the static scheduler of the \textit{PLASMA} library \cite{buttari2009class, agullo2009plasma} to implement a sparsity-aware tile Cholesky factorization on x86 systems equipped with GPU hardware accelerators. 
By relying on tile algorithms to achieve fine-grained parallelism, \emph{sTiles} computes on dense clusters within block arrowhead matrices, while balancing the number of extra floating-point operations and the arithmetic intensity with the tile size. The arrowhead matrix pattern does not expose enough parallelism due to the data dependencies. Using the static scheduler, we enforce a left-looking variant of the Cholesky factorization, which enables \emph{sTiles} to use a tree reduction for concurrent updates and ultimately increase the degree of parallelism. This technique alleviates the performance bottleneck highlighted in dense-to-sparse solver transitions in the context of low-rank matrix approximations for data sparse problems \cite{cao2022framework}. \emph{sTiles} has a direct impact on effectively handling the Gaussian Markov Random Fields \cite{rue2005gaussian}, where their inverse covariance matrices (precision matrices) often display densely connected clusters, thereby enhancing the overall performance of large-scale Bayesian inference applications.

Our contributions in this paper are multifaceted, as follows:
\begin{enumerate}
    \item \textbf{Comprehensive analysis for preprocessing phase}: We conduct a comprehensive analysis of permutation techniques for sparse Cholesky factorization, focusing on Reverse Cuthill-McKee (RCM) \cite{george1981computer}, Approximate Minimum Degree (AMD) \cite{amestoy2004algorithm}, and Nested Dissection (ND) \cite{george1973nested}, to evaluate their effectiveness for block arrowhead matrices.

    \item \textbf{Development of \textit{sTiles} framework}: We introduce \textit{sTiles}, specifically designed for sparse structured matrices with a focus on block arrowhead matrices. Unlike previous works limited to block tridiagonal matrices and spatiotemporal applications \cite{gaedke2024integrated}, \textit{sTiles} can extend the scope to full bandwidth (fully dense) matrices if required by the application.

    \item \textbf{Parallel capabilities in \textit{sTiles}}: We enhance \textit{sTiles} with new features tailored to block arrowhead matrices, including a tree reduction technique, supporting GPU acceleration, and launching multiple concurrent Cholesky factorizations.

    \item \textbf{Performance comparison against state-of-the-art solvers}: We assess the proposed sparsity-aware Cholesky factorization against \textit{CHOLMOD} \cite{davis2005cholmod}, \textit{SymPACK} \cite{jacquelinsympack}, \textit{MUMPS} \cite{amestoy2000mumps}, and \textit{PARDISO} \cite{schenk2001pardiso}, achieving up to 8.41X/9.34X/5.07X/11.08X speedups, respectively. On GPUs, \emph{sTiles} attains a 5X speedup against its CPU-only version. %These comparisons demonstrate significant improvements in efficiency and scalability.
\end{enumerate} %

These contributions collectively advance the support for efficient factorization of block arrowhead matrices, particularly within the context of large-scale Bayesian inference applications.

The remainder of the paper is as follows. Section \ref{sec1} provides context on tile Cholesky factorization. Section \ref{sec2} focuses on the specific challenges of block arrowhead matrices, discussing permutation techniques and the necessary development of a static scheduling algorithm tailored to the Cholesky factorization for sparse structured matrices. In Section \ref{sec3}, we explore additional parallelization strategies, including tree reduction, concurrent factorizations, and the support of GPU acceleration for performance improvement. Finally, the results of extensive performance evaluations are presented in Section \ref{sec4} and we conclude in Section \ref{sec5}.

\section{Background on Dense-Sparse Cholesky Factorization} \label{sec1}

The aim of the Cholesky factorization is to decompose a symmetric, positive-definite \( N \times N \) matrix \( A \) into the form \( A = LL^T \), where \( L \) is a lower triangular matrix. The sparse Cholesky computation is typically carried out in three steps:

\begin{enumerate}
    \item \textbf{Heuristic Reordering}: The rows and columns of \( A \) are reordered to minimize fill-in in the factor matrix \( L \). This step enhances the efficiency of the factorization by reducing the number of nonzero elements.
    \item \textbf{Symbolic Factorization}: This step involves determining the nonzero structure of \( L \) based on the reordering. It is a symbolic computation that identifies where the nonzero elements will be located, allowing for the allocation of storage for \( L \).
    \item \textbf{Numerical Factorization}: In this final step, the actual numerical values of the nonzero elements in \( L \) are computed. This is the most computationally intensive part of the process.
\end{enumerate}

While traditional sparse Cholesky factorization is effective and fast for non-structured matrices with scattered nonzero elements, it fails to exploit block structures when applied to large matrices with densely connected regions. An example of this is arrowhead matrices, where the structure consists of a dense core and sparse outer regions (see Figure \ref{patternsAHM}). To deploy high performance matrix computations on such  sparse structured matrix, we bring in the concept of tile algorithms for fine-grained computations from the well-established dense linear algebra community~\cite{buttari2009class, SOlQuintana, agullo2009plasma}. This provides flexibility to directly operate on nonzero tiles, while benefiting from task orchestration mapped onto parallel processing units via a runtime system. The sparse tile Cholesky factorization can then be translated into a directed acyclic graph (DAG), where nodes are computational tasks and edges represent data dependencies between them.

Different forms of Cholesky factorization, specifically column-oriented algorithms, both right-looking (supernodal) and left-looking variants (multifrontal), are particularly effective for handling sparse matrix computations \cite{heath1997parallel}. 
Given the presence of zero tiles, we present the algorithm for the sparse tile Cholesky factorization (left-looking variant) as detailed in Algorithm \ref{algo:left-looking-cholesky}. The algorithm relies on four basic operations implemented by computational kernels: \textbf{POTRF} carries out the Cholesky factorization of a diagonal tile, \textbf{SYRK} performs a symmetric rank-k update on a diagonal tile, \textbf{TRSM} applies an update to an off-diagonal tile by performing a triangular solve, and \textbf{GEMM} updates an off-diagonal tile through a matrix-matrix multiplication.

Here, \( m \), \( k \), and \( n \) represent indices corresponding to different tiles of the matrix. We define $\text{neighbors}(k)$ for tiles such that $m \in \text{neighbors}(k)$ if $A_{mk}$ or $A_{km}$ is a nonzero tile.
{\footnotesize
\begin{algorithm}
    \caption{Pseudocode of the sparse tile Cholesky factorization (left-looking version)}
    \label{algo:left-looking-cholesky}
    \begin{algorithmic}[1] % Add line numbers
        \For{\(k \gets 0\) \textbf{to} Tiles\(-1\)}
            \ForAll{\(n \in \text{neighbors}(k)\)}
                \State \(A_{kk} \gets \texttt{SYRK}(A_{nk}^T, A_{kk})\)
            \EndFor
            \State \(A_{kk} \gets \texttt{POTRF}(A_{kk})\)
            \For{\(m \gets k+1\) \textbf{to} Tiles\(-1\)}
                \ForAll{\(n \in \text{neighbors}(k) \cap \text{neighbors}(m)\) \textbf{and} \(n < k-1\)}
                    \State \(A_{km} \gets \texttt{GEMM}(A_{nk}^T, A_{mn}, A_{km})\)
                \EndFor
                \If{\(m \in \text{neighbors}(k)\)}
                    \State \(A_{km} \gets \texttt{TRSM}(A_{kk}, A_{km})\)
                \EndIf
            \EndFor
        \EndFor
    \end{algorithmic}
\end{algorithm}}

\begin{figure}
    \begin{subfigure}[b]{0.6\columnwidth}
        \centering
        \includegraphics[width=\columnwidth]{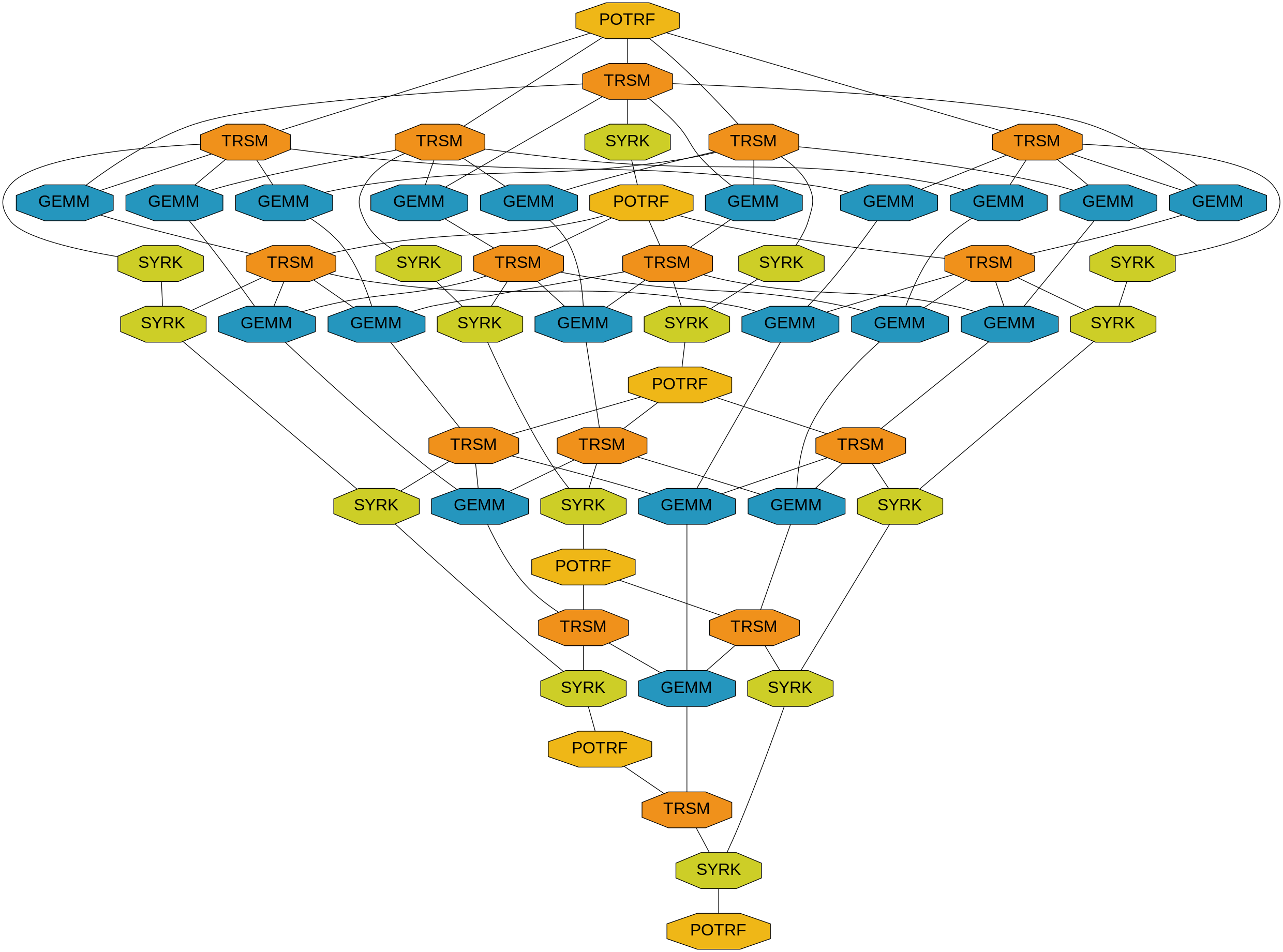}
    \end{subfigure}
    \begin{subfigure}[b]{0.16\columnwidth}
        \centering
        \includegraphics[width=\columnwidth]{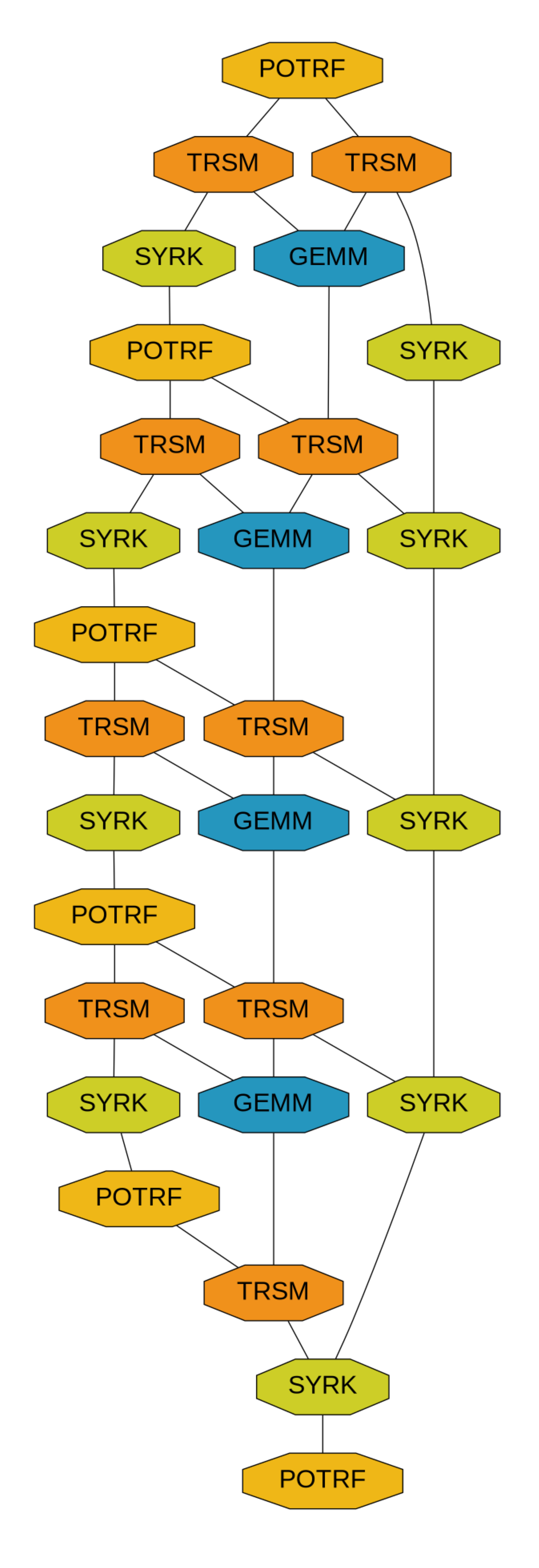}
    \end{subfigure}
    \begin{subfigure}[b]{0.2\columnwidth}
        \centering
        \includegraphics[width=\columnwidth]{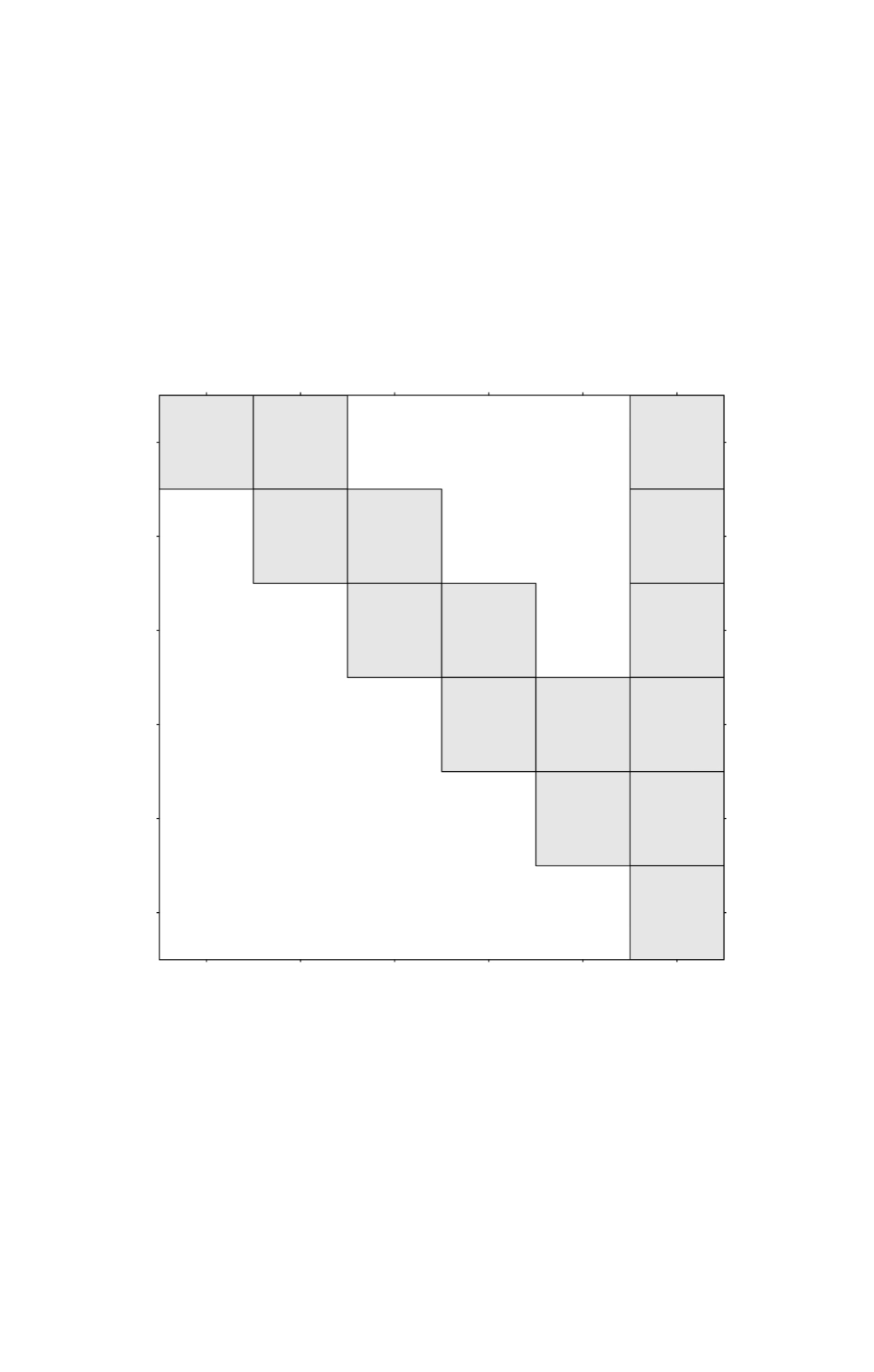}
    \end{subfigure}
    \caption{DAG representations of task dependencies for Cholesky factorization on dense and arrowhead tiled matrices (6x6 tile configuration).}
    \label{dag-treesss}
\end{figure}
For example, Figure \ref{dag-treesss} presents the Directed Acyclic Graph (DAG) representations of the tasks involved in the Cholesky factorization process for a tile  dense matrix (left) and a tile sparse arrowhead structured matrix (right). Each node in the DAG represents a computational task (e.g., POTRF (yellow), TRSM (orange), SYRK (green), GEMM (blue)), and the directed edges between nodes illustrate the dependencies between these tasks. The width of the dense matrix’s DAG shows the maximum degree of parallelism while the height depicts the length of the critical path.
In contrast, the DAG for the arrowhead matrix is much thinner, indicating challenges for parallelization due to the limited opportunities for task concurrency. 

The right-looking variant should be privileged in presence of high concurrency since many updates on the trailing submatrix can occur simultaneously. However, due to the arrowhead matrix structure, the degree of parallelism is limited. With a left-looking variant, the GEMM operations behave as an accumulator, which provides an opportunity to further expose parallelism that will be discussed later in the paper.
Consequently, our focus is on the left-looking variant of the algorithm driven by a customized static scheduler to enforce a left-looking traversal of the thin DAG. 

\section{Arrowhead Sparse Matrix Factorization}  \label{sec2}

\subsection{Permutation Techniques}

Exploiting the structure of arrowhead matrices is paramount for efficiency. Investing time in the preprocessing stage is particularly important when multiple Cholesky decompositions are required based on the same structure. We assume the initial configuration of the arrowhead shape points to the bottom right corner, as it decreases the fill-in during the Cholesky decomposition, resulting in more efficient storage and faster computations, see Figure \ref{patternsAHM}. We aim to make the arrowhead shape thinner, as this decreases the number of dependent computational routines.

When working with block arrowhead matrices, two types of permutations are considered: a complete permutation reorders the entire matrix, including both the diagonal and arrowhead region, while a partial permutation focuses on reordering a submatrix, typically the diagonal part, leaving the dense arrowhead region untouched.

Given that arrowhead matrices are still sparse, considering ordering techniques can further optimize the structure. In this context, we explore three primary techniques: Reverse Cuthill-McKee (RCM), Approximate Minimum Degree (AMD) \cite{amestoy2004algorithm}, and Nested Dissection (ND). 

\begin{figure}
    \centering
        \includegraphics[width=0.7\columnwidth]{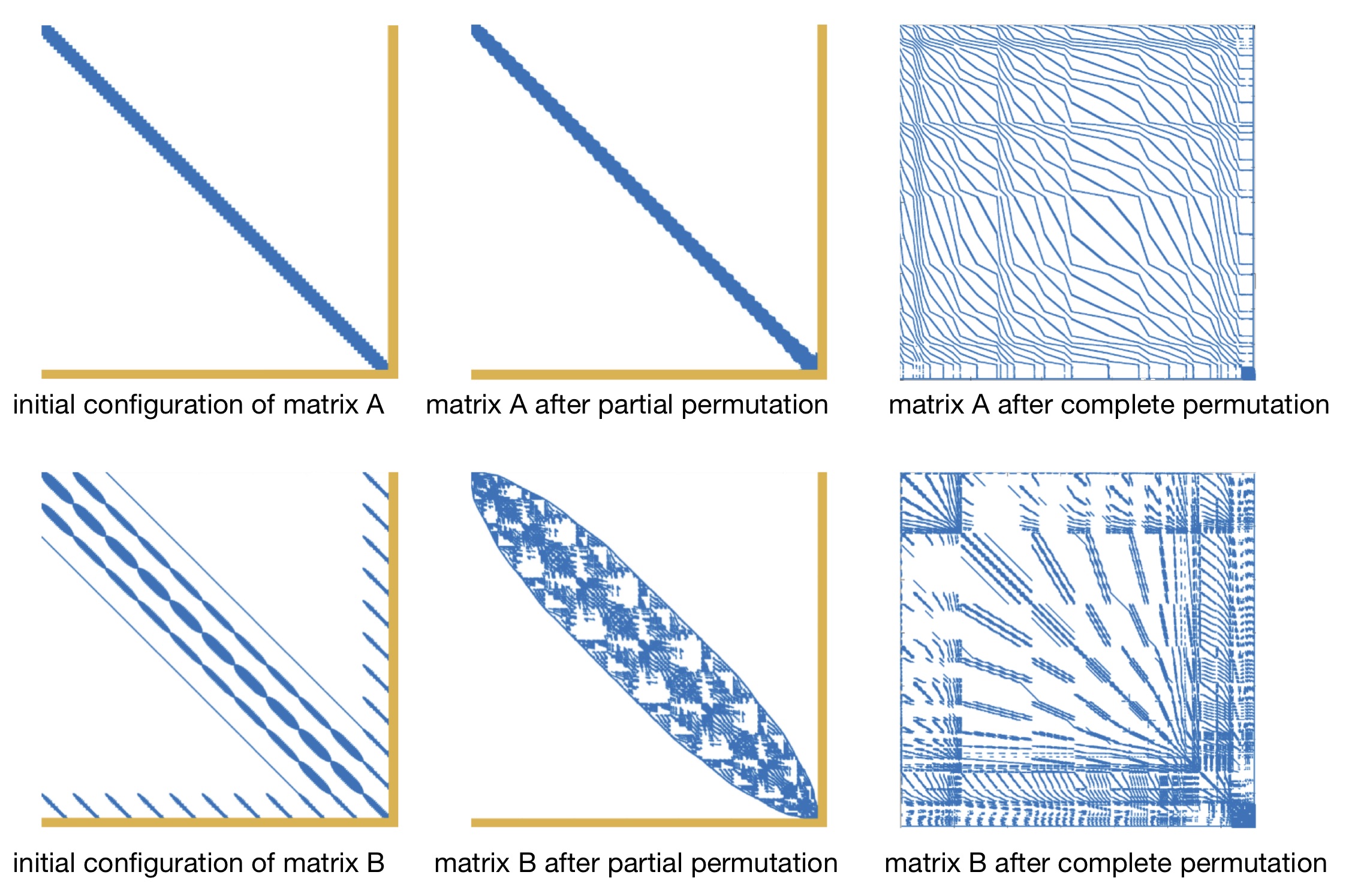}
        \caption{Visualization of matrix permutations using RCM: highlighting unaltered segments (orange part) in partial permutations.}
        \label{RCMpatterns}
\end{figure}

\noindent 1. \textbf{RCM:} we explore the RCM algorithm, commonly used to reduce the bandwidth of sparse matrices by reordering nodes to minimize their distance from the diagonal. Applying RCM to our arrowhead matrices, particularly preserving the ``arrowhead region" untouched, resulted in better permutation outcomes. Excluding the arrowhead regions (in orange in Figure~\ref{RCMpatterns}) decreased fill-in significantly and maintained a more orderly matrix structure, contrasting sharply with the disordered outcomes when the orange part was included. Matrix \textit{A}'s initial diagonal structure indicated that little to no permutation was necessary, with no fill-ins required across different configurations. Conversely, Matrix B saw a significant decrease in fill-ins, reducing by approximately 32.71$\%$ in the second configuration (partial RCM permutation) compared to the initial configuration. This illustrates the effectiveness of selectively applying permutations, particularly avoiding the orange arrowhead regions (dense arrowhead).% enhancing matrix performance without negatively impacting computational efficiency.

\noindent \textbf{2. AMD:} AMD selects nodes of the least degree for elimination to reduce fill-in, implicitly representing new edges as cliques. After selecting and eliminating a node, AMD updates the degrees of the neighboring nodes to reflect the new connections formed by the elimination process. 
%This update is done approximately, rather than exactly, to save computational time. 
Although it is not the best choice for arrowhead matrices due to their unique structured patterns, AMD remains an option for matrices with less predictable or irregular patterns, including those arising from poor quality meshes, where reducing computational overhead and storage requirements is crucial \cite{kilic2012effect}. 

\noindent \textbf{3. ND:} The key idea of nested dissection algorithm is to recursively partition the graph of the matrix into smaller subgraphs by selecting a vertex separator, which is a set of vertices whose removal splits the graph into approximately equal-sized, disconnected subgraphs. The size of the separator plays a crucial role in the efficiency of the algorithm; smaller separators generally lead to less fill-in and more efficient decompositions. The chosen vertices are eliminated last, which helps to maintain sparsity in the resulting matrix. 

\begin{figure}
    \centering
        \includegraphics[width=0.7\columnwidth]{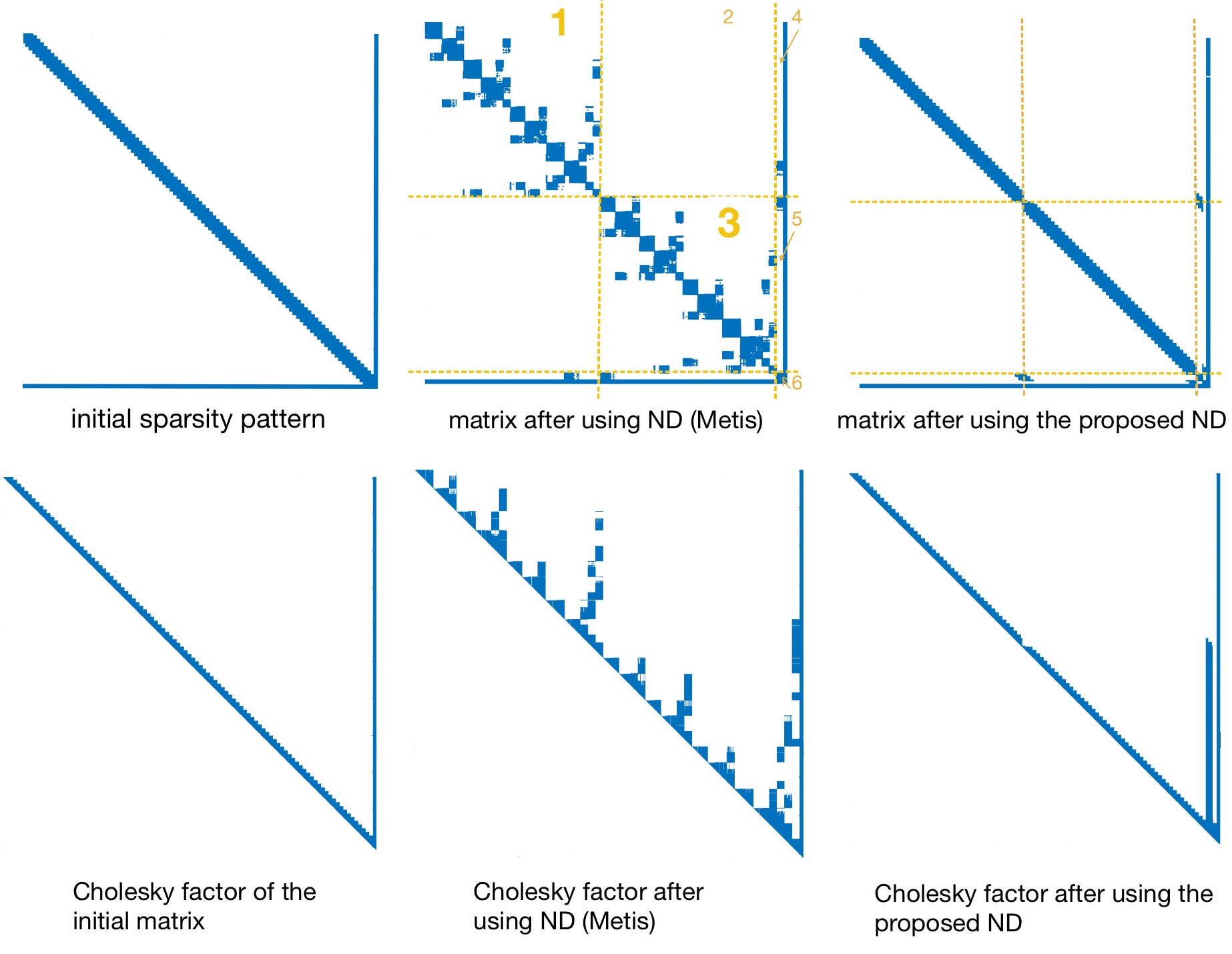}
        \caption{Comparison of sparsity patterns and Cholesky factors: initial matrix, ND (METIS), and proposed ND.}
        \label{NDpatterns}
\end{figure}

ND appeals because partitions can be processed in parallel. This means that computations can run concurrently without dependencies between the partitions, allowing efficient use of multi-core processors. In Figure \ref{NDpatterns}, we can see that after applying ND using METIS \cite{karypis1997metis}, the matrix is divided into distinct blocks (top middle), which can be processed independently (block 1 and 3). However, the resulting partitions are not always optimal for certain matrix structures, such as arrowhead matrices. The same figure (top middle) illustrates this issue, where the reordering results in a dispersed pattern does not preserve the arrowhead structure.

We propose an adaptable nested dissection method that addresses these challenges. Our approach involves:
\begin{enumerate}
    \item Computing the bandwidth: we first compute the bandwidth of the matrix. The size of the separator is equal to the bandwidth and the number of columns in the arrowhead shape.
    \item Adjusting the separator position: the separator size is moved towards the end of the matrix. This adjustment preserves the shape of the matrix and is more effective than the default method used by METIS for arrowhead matrices.
\end{enumerate}
        
Figure \ref{NDpatterns} illustrates that the matrix reordered using the proposed ND method maintains a more structured pattern compared to the METIS approach. The Cholesky factor of the matrix after applying the proposed ND method also demonstrates a sparser fill-in, indicating more efficient decomposition and better preservation of sparsity structure. Without any ordering, the matrix retains a thin arrowhead structure, which results in high computational dependency and limited scalability. When using the METIS ordering, the reordering is performed in a generic manner. Even with partial ND using METIS, the structure is not preserved, leading to a dispersed pattern. While this approach introduces parallelism, it also results in more fill-ins compared to having no ordering. In contrast, the proposed ND method preserves the original structure more effectively, albeit with a slightly thicker arrow shape in the second partition. This adjustment introduces a minor imbalance in task distribution between partition 1 and partition 2 but enhances parallelism in the computational process.

An important consideration for applying ordering techniques to arrowhead matrices is to first understand the pattern of the given matrix and choose the appropriate method accordingly. Given the variety of possible structures, no single method is best for all arrowhead matrices. However, understanding the characteristics of each method allows us to make an informed choice:

\textit{Partial RCM is preferred for reducing bandwidth while preserving matrix structure. Adaptable ND is recommended when enough cores ($\ge$ 6) are available and the separator is not large. AMD works well for matrices with irregular patterns, while ND (METIS) is a generic option, though it often increases fill-ins. For any technique, the number of fill-ins is evaluated before and after the ordering; if there is no improvement, the method is not used. 
}

\subsection{Compressed Tile Storage Format (CTSF)}

\begin{figure}%[!t]
    \centering
    \begin{subfigure}[b]{0.35\columnwidth} % Adjust width to fit content properly
        \centering
        \includegraphics[width=\columnwidth]{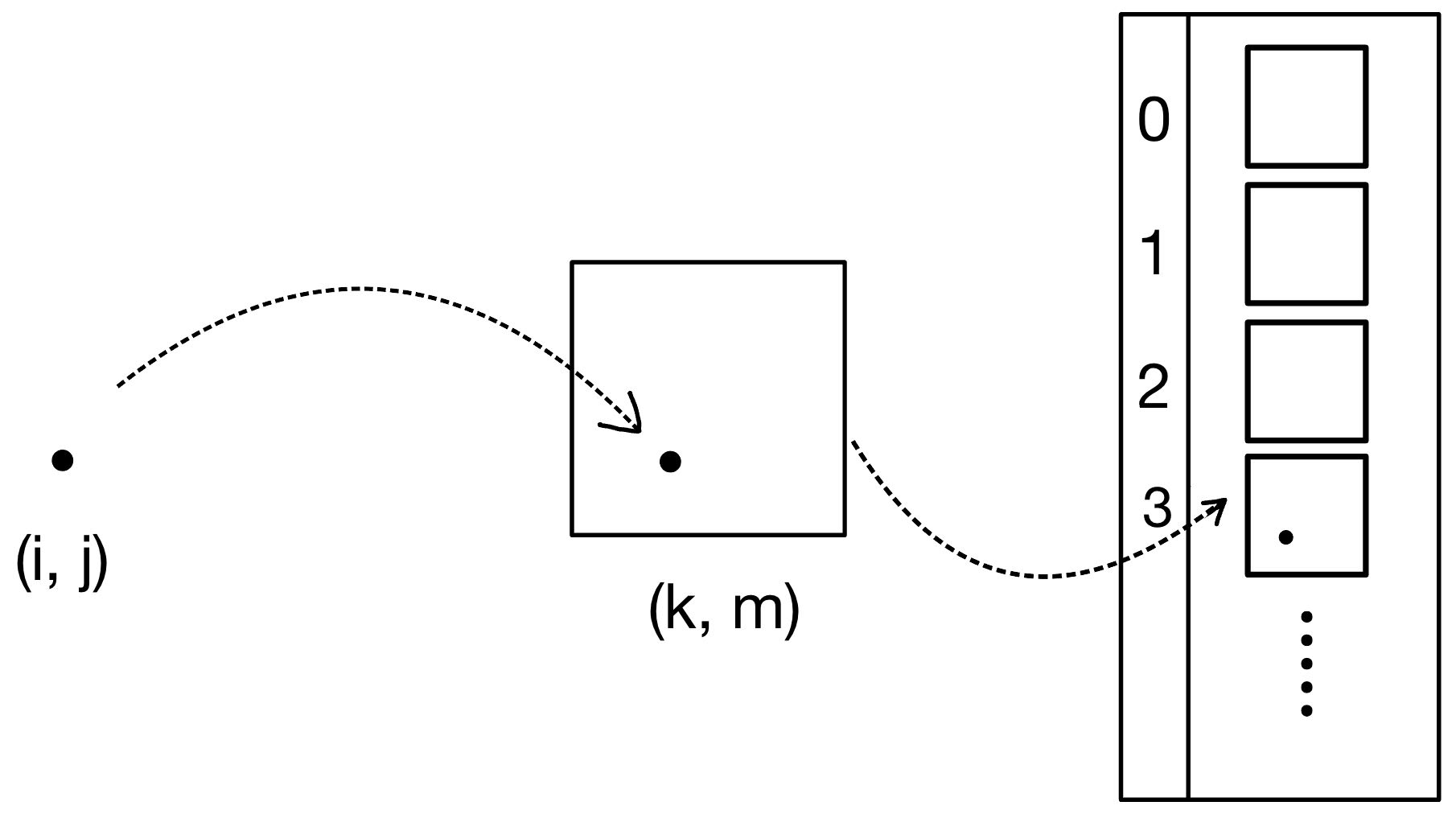} % Use \columnwidth to fit within subfigure
        %\caption{Mapping}
        \label{fig:mapping}
    \end{subfigure}
    \begin{subfigure}[b]{0.2\columnwidth} % Adjust width similarly
        \centering
        \includegraphics[width=\columnwidth]{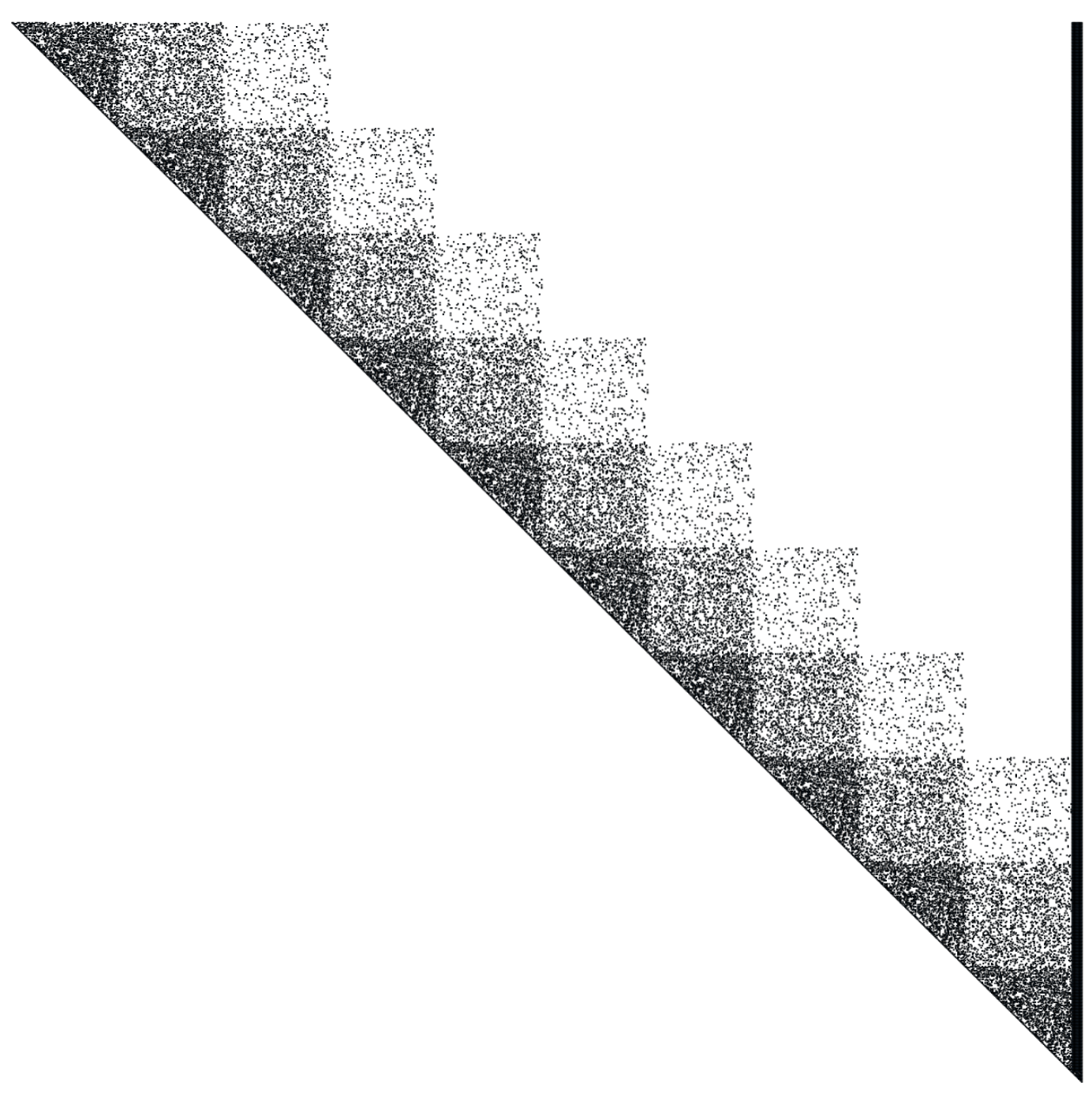}
        %\caption{S Pattern}
        \label{fig:spattern}
    \end{subfigure}
    \begin{subfigure}[b]{0.2\columnwidth}
        \centering
        \includegraphics[width=\columnwidth]{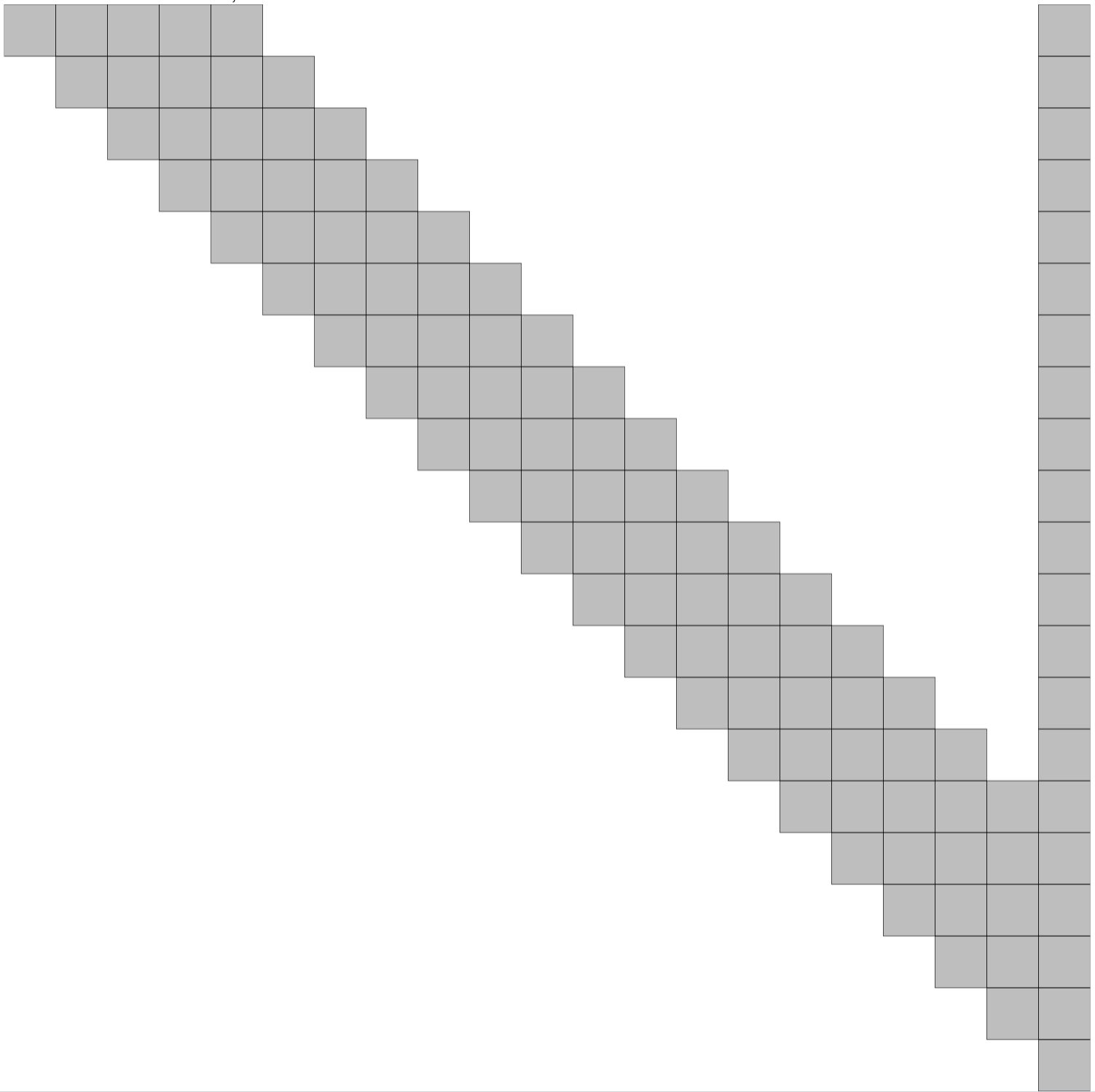}
        %\caption{D Pattern}
        \label{fig:dpattern}
    \end{subfigure}
    \caption{Mapping of elements from a sparse matrix to a sparse tiled matrix.}
    \label{fig:mappingtotiles}
\end{figure}

The Compressed Tile Storage Format (CTSF) is introduced as an efficient method to handle sparse matrices containing dense connected clusters by transforming them into a tile data format. This approach divides the sparse matrix into smaller tiles, each of a fixed size \(N_T \times M_T\), where the tile size is independent of the block arrowhead shape. Each element \((i, j)\) in the sparse matrix is mapped to a corresponding tile \((k, m)\), which is allocated only when an element is mapped to it (initially, all elements in the tile are set to zero). These tiles are then organized into contiguous memory blocks, facilitating more efficient storage and computation. The sparse elements are read in a Compressed Sparse Column (CSC) format. Figure \ref{fig:mappingtotiles} illustrates the mapping of elements from a sparse matrix to a sparse tiled matrix.

The process of mapping elements to tiles may result in a structure that does not strictly follow an arrowhead shape. Nonetheless, the factorization process can still proceed. In such cases, it may be advantageous to apply an additional layer of tile ordering (permutation). However, this paper primarily focuses on assuming a block arrowhead matrix.

Moreover, working with tiles as a basic block offers several advantages. Tile algorithms offer granularity choices for achieving high efficiency on various parallel multicore systems, while facilitating task scheduling based on dataflow graph using a runtime system~\cite{kurzak2006implementing, buttari2009class,SOlQuintana}.

\subsection{Sparse Cholesky Algorithm using Static Scheduling}

We describe the algorithm used for the static version of sparse matrix factorization, specifically tailored for arrowhead matrices, as explained in Algorithm~\ref{algo:static}. Our approach leverages the principles outlined in the static pipeline scheduling for dense matrix factorizations, originally discussed in \cite{kurzak2010scheduling}. This technique is simple yet effective, providing good data locality and load balance for regular computations, such as dense matrix operations.

The static pipeline scheduling approach identifies each task by an $\{m, n, k\}$ triple, which defines the operation type and tile location \cite{kurzak2010scheduling}. Each thread traverses its tasks using a formula based on its ID and the total number of cores. Task dependencies are tracked with a global progress table. Before executing a task, each thread checks this table for dependencies, stalling if necessary. Upon task completion, the thread updates the table. The table is volatile, with updates managed by writing to an element and stalls handled with busy waits.

In our approach for sparse matrix factorization of arrowhead matrices, we adapt the static pipeline scheduling principles. After applying permutation and symbolic factorization to optimize the sparsity pattern and reduce fill-ins, the algorithm proceeds with numerical factorization using a customized sparsity-aware static scheduling. Each thread executes a predefined set of tasks based on its assigned portion of the matrix, ensuring parallel execution and efficient use of computational resources.

Given a matrix \textit{A} in CTSF, we initialize a global progress table (\texttt{core\_progress}) and perform Cholesky factorization on the upper part of the matrix as follows. Each thread has its own version of a Task Assignment Table (TAT), filled during a preprocessing stage. The TAT contains the tasks each core needs to perform. 

\begin{algorithm}
    \caption{Asynchronous Task-based Sparse Cholesky Factorization.}
    \begin{algorithmic}
    \For{\texttt{each thread ID}}
        \For{\texttt{each task i assigned to thread ID}}
            \State \texttt{m = TAT[i, 0]}
            \State \texttt{k = TAT[i, 1]}
            \State \texttt{n = TAT[i, 2]}
            \State \texttt{task\_type = TAT[i, 3]}
            \If{\texttt{task\_type = 1}}
                \State $A_{kk} \gets \texttt{POTRF}(A_{kk})$
                \State \texttt{Set core\_progress[k,k] = 1}
            \ElsIf{\texttt{task\_type = 2}}
                \While{\texttt{core\_progress[n,k] $\neq$ 1}}
                    \State \texttt{/* Wait */}
                \EndWhile
                \State $A_{kk} \gets \texttt{SYRK}(A_{nk}^T, A_{kk})$
            \ElsIf{\texttt{task\_type = 3}}
                \While{\texttt{core\_progress[k,k] $\neq$ 1}}
                    \State \texttt{/* Wait */}
                \EndWhile
                \State $A_{km} \gets \texttt{TRSM}(A_{kk}, A_{km})$
                \State \texttt{Set core\_progress[m,k] = 1}
            \ElsIf{\texttt{task\_type = 4}}
                \While{\texttt{core\_progress[k,n] $\neq$ 1}}
                    \State \texttt{/* Wait */}
                \EndWhile
                \While{\texttt{core\_progress[m,n] $\neq$ 1}}
                    \State \texttt{/* Wait */}
                \EndWhile
                \State $A_{km} \gets \texttt{GEMM}(A_{nk}^T, A_{mn}, A_{km})$
            \EndIf
        \EndFor
    \EndFor
    \end{algorithmic}
    \label{algo:static}
\end{algorithm}

This static scheduling approach ensures efficient parallel computation by leveraging predefined task assignments and managing dependencies through the global progress table. The code implements a left-looking version of the factorization for sparse matrices, where work is distributed by rows of sparse tiles and the factorization steps are pipelined. In this approach, once a thread completes its assigned tasks for step \(i\), it immediately begins the factorization of the panel in step \(i+1\) once the corresponding data dependencies are satisfied. Following threads then proceed to update operations for step \(i+1\) before moving on to the panel in step \(i+2\). This pipelined execution (or \textit{lookahead}) ensures continuous progress and efficient use of computational resources by overlapping computation and communication phases.

\section{Parallel Optimization Techniques and GPU Acceleration}  \label{sec3}

The Directed Acyclic Graph (DAG) representation of tasks in sparse Cholesky factorization for an arrowhead matrix reveals a thinner and more streamlined structure compared to general matrices, as depicted in Figure \ref{dag-treesss}. This streamlined structure results in limited parallelism due to the inherent dependencies in the computation. Consequently, adding more computational cores beyond a certain point does not significantly enhance performance. The dependency chains create bottlenecks, preventing further speedup despite the availability of additional cores. Tree reduction is one technique to mitigate these limitations by enhancing parallel efficiency. This section details tree reduction and its role in optimizing sparse Cholesky factorization with arrowhead structure. Additionally, Section~\ref{GPUdetails} covers GPU optimization in \textit{sTiles}, with further strategies in Appendix A.

\subsection{Tree Reduction}

\begin{figure}
    \centering
    \begin{subfigure}[b]{\columnwidth}
        \centering
        \includegraphics[width=0.8\columnwidth]{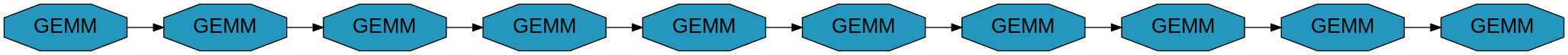}
    \end{subfigure}
    \\[0.05cm]
    \rule{0.5\columnwidth}{0.1pt}
    \\[0.1cm]
    \begin{subfigure}[b]{\columnwidth}
        \centering
        \includegraphics[width=0.8\columnwidth]{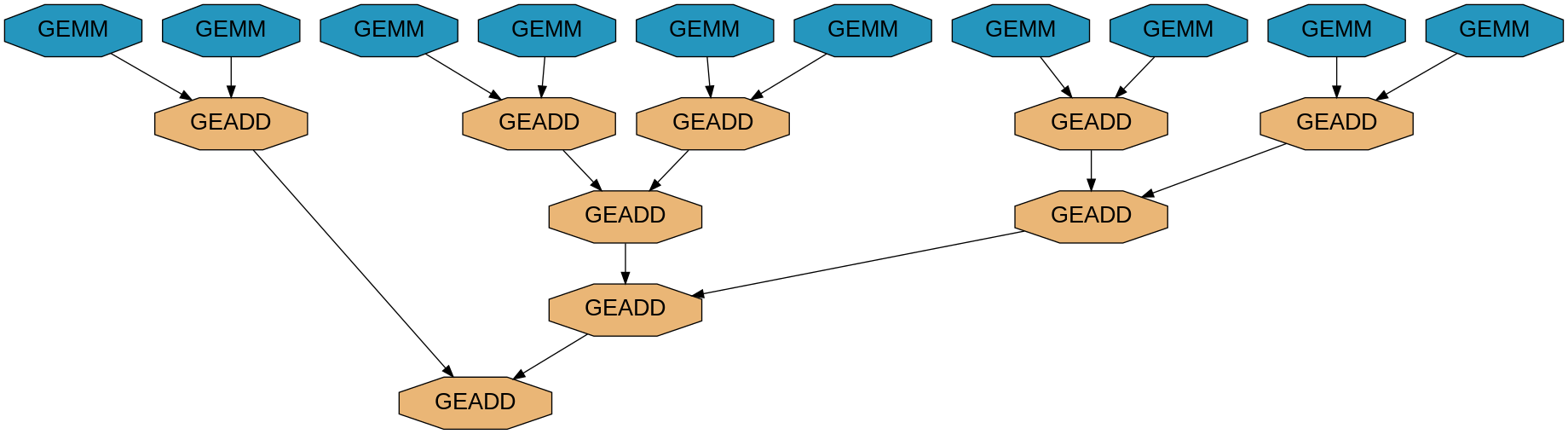}
    \end{subfigure}
    \caption{The first subplot depicts the sequential execution of GEMMs, while the second subplot shows the tree reduction approach using Generalized Additions (GEADDs).}
    \label{SPGEMMS}
\end{figure}

\textbf{Introduction:} Tree reduction restructures computation into a hierarchical tree to enable efficient parallel execution by aggregating intermediate results in a structured manner, while favoring local computations/reductions to mitigate data movement overheads. In Cholesky factorization, particularly with thick arrowhead matrices, GEMM (General Matrix-Matrix Multiplication) and SYRK (Symmetric Rank-K update) operations accumulate successive inter-dependent updates. Tree reduction organizes these accumulations hierarchically to allow efficient aggregation of intermediate results, while further exposing parallelism. Figure \ref{SPGEMMS} illustrates the sequential execution of GEMMs and their organization using tree reduction through Generalized Additions (GEADDs).

The execution times for running \( k \) GEMMs or SYRKs sequentially highlight the growing computational bottlenecks. Table \ref{SEQtime} presents these times for varying \( k \), showing a near-linear increase in execution time. This pattern underscores the need for parallel strategies to mitigate the bottlenecks related to sequential execution.

\begin{table}%[h]
\centering
\begin{tabular}{c|c|c}
\textbf{$k$} & \textbf{GEMMs Time (s)} & \textbf{SYRKs Time (s)} \\ \hline
1,000        & 0.091                   & 0.088                   \\ \hline
5,000        & 0.447                   & 0.430                   \\ \hline
10,000       & 0.916                   & 0.865                   \\ \hline
50,000       & 4.547                   & 4.342                   \\ 
\end{tabular}
\caption{Execution times for running \( k \) GEMMs or SYRKs sequentially.}
\label{SEQtime}
\end{table}

\begin{figure}%[htbp]
    \centering
    \includegraphics[width=0.8\columnwidth]{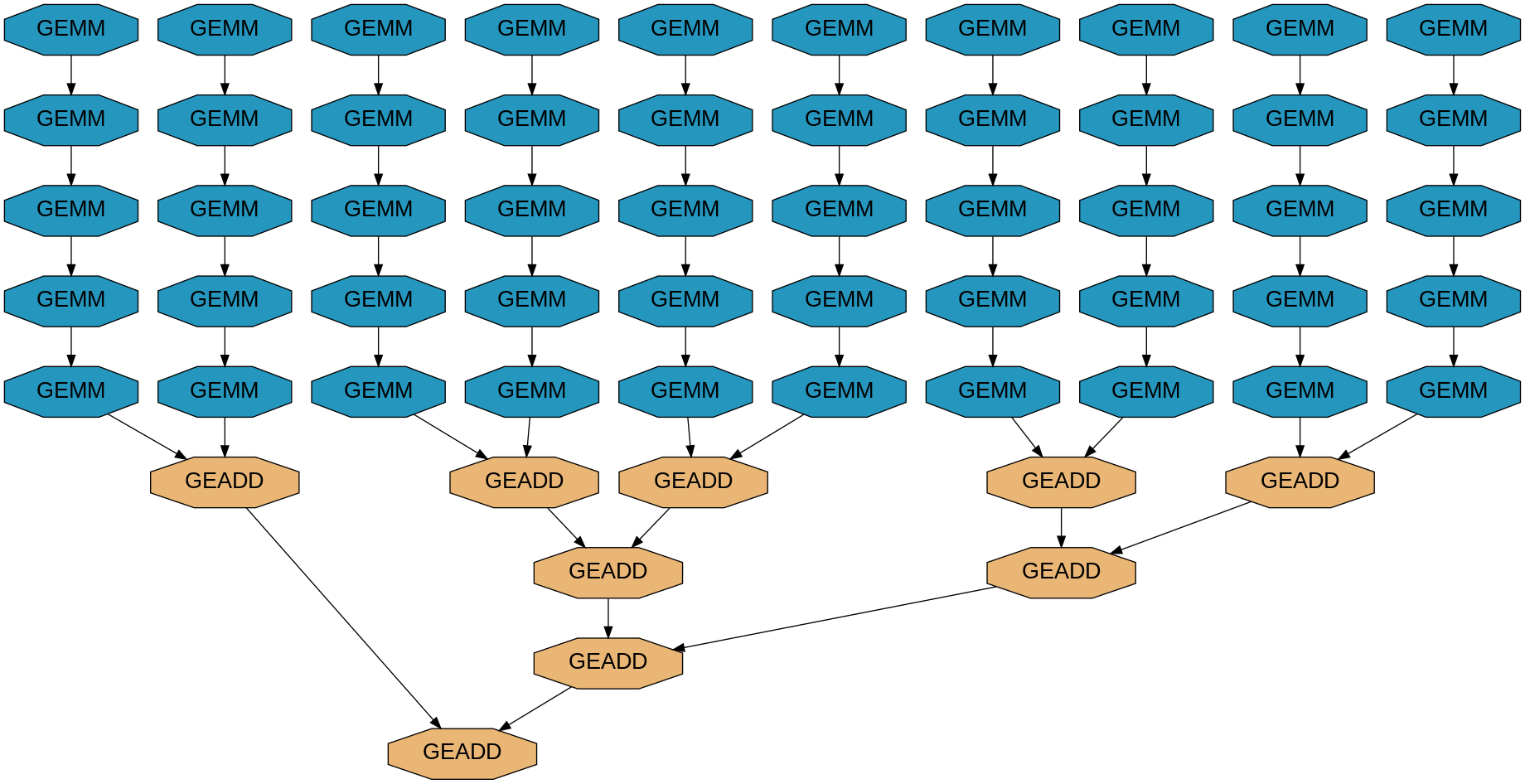}
    \caption{Tree reduction approach preceded by sequential GEMMs, where the number of GEMM executions corresponds to the number of available cores.}
    \label{PA1SA2}
\end{figure}

\textbf{Algorithm for Tree Reduction:} The tree reduction algorithm employs a binary tree structure. Each core is assigned a subset of GEMMs, and the results are progressively aggregated using GEADD operations. Memory is allocated for intermediate results during tree reduction, and the final result is computed by combining all intermediate results hierarchically, see an illustration in Figure \ref{PA1SA2}. The details are outlined in Algorithm \ref{alg:tree_reduction}.

\begin{algorithm}
\caption{Tree Reduction for Parallel Cholesky Factorization}
\label{alg:tree_reduction}
\begin{algorithmic}
\State \textbf{Input:} Tiled matrix $A$, thread count: $num\_of\_cores$, GEMM index range $\{start\_range, end\_range\}$
\State \textbf{Output:} Reduced result $T_{final}$

\State Initialize an array of tiles, \texttt{T[ID]} for partial results, one per core
\State Initialize \texttt{core\_progress[ID]} $\gets 0$ for all threads
\For{\texttt{each thread ID} $= 0$ \textbf{to} \texttt{num\_of\_cores - 1}} \Comment{Parallel region}
    \For{\texttt{each GEMM with indices $\{k, m, n\}$}}
        \State \texttt{i $\gets$ determine\_mapping\_index($k, m, n$)}
        \If{\texttt{start\_range[ID]} $\leq$ \texttt{i} $<$ \texttt{end\_range[ID]}}
            \State Perform GEMM for assigned range:
            \State $T_\texttt{ID} \gets T_\texttt{ID} + $ \texttt{DGEMM($A_{kk}, A_{mk}$)} 
        \EndIf
    \EndFor
    \State \texttt{core\_progress[ID]} $\gets$ \texttt{1} \Comment{Mark thread completion}
\EndFor
\State \textbf{Synchronize all threads}
\State Perform hierarchical reduction of \texttt{T[ID]} using GEADD to compute $T_{final}$
\end{algorithmic}
\end{algorithm}

\begin{figure}
    \centering
    \begin{minipage}[b]{\columnwidth}
        \centering
        \includegraphics[width=0.8\columnwidth]{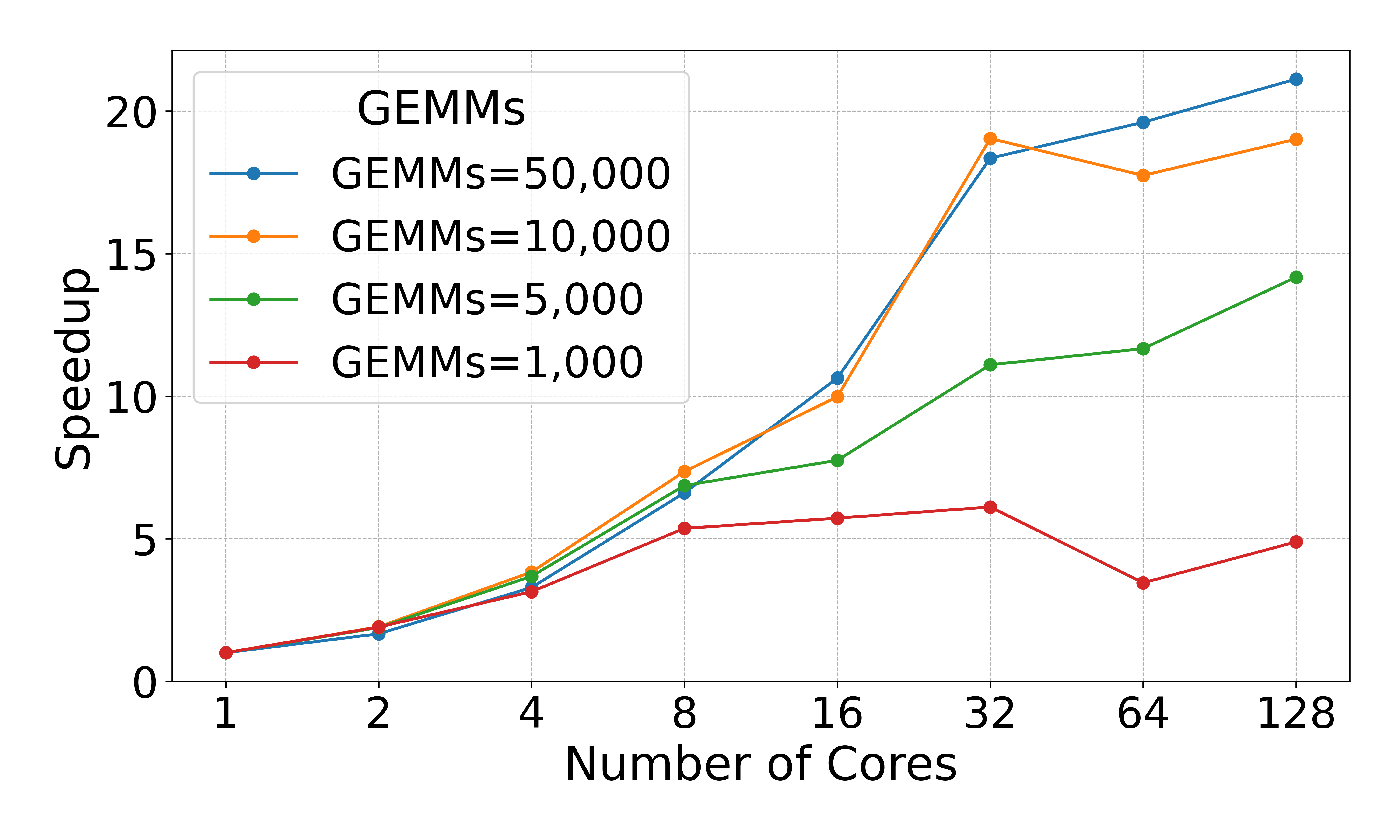}
        \caption{Speedup achieved using tree reduction for different core counts, compared to sequential execution.}
        \label{PA1}
    \end{minipage}
    \hspace{0.05\columnwidth}
    \begin{minipage}[b]{\columnwidth}
        \centering
        \includegraphics[width=0.8\columnwidth]{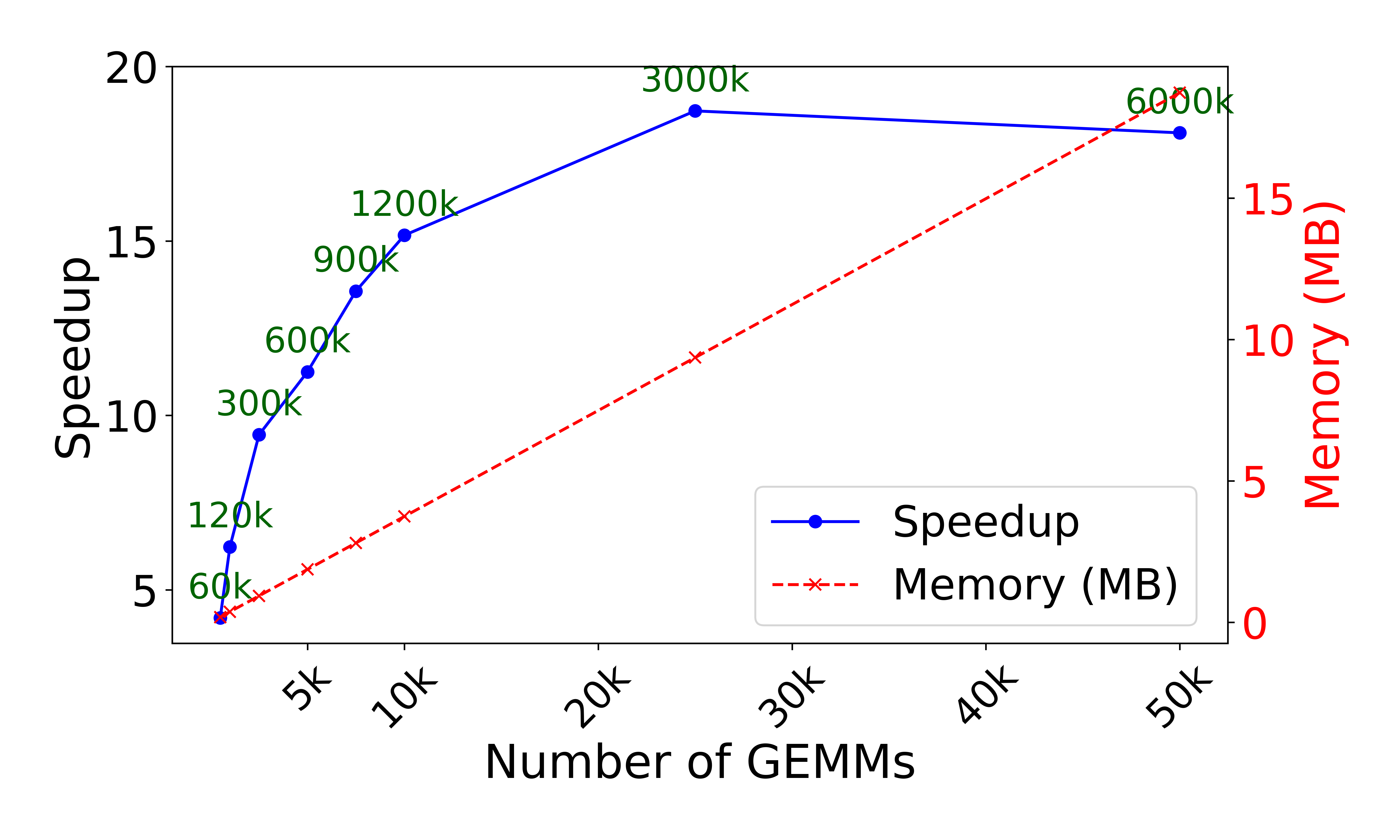}
        \caption{Speedup, memory usage, and matrix size (green) vs number of GEMMs computed using tree reduction with 32 cores.}
        \label{PA2}
    \end{minipage}
\end{figure}

\textbf{Performance Analysis:} Using tree reduction, we achieve significant improvements in speedup and memory efficiency. Figure \ref{PA1} illustrates the speedup achieved by applying tree reduction across different core counts. For larger GEMMs, the speedup stabilizes as the number of cores increases, with improvements of up to 20X compared to sequential execution. Figure \ref{PA2} shows the relationship between speedup, memory usage, and matrix size for 32 cores, highlighting the balance between computation and memory footprint.

\textit{Based on this analysis, sTiles adopts the tree reduction strategy with the condition that the number of GEMMs is at least twice the number of available cores.}

\subsection{Leveraging GPUs in \textit{sTiles}: Performance Gains and Considerations} \label{GPUdetails}

The use of dense tiles in the \textit{sTiles} framework naturally motivates the integration of GPU acceleration to further enhance computational performance by launching \texttt{cuBLAS}/\texttt{cuSOLVER} kernels to the GPU. However, several challenges arise when adapting the \textit{sTiles} approach for GPU architectures, requiring careful consideration of both the problem size and hardware limitations. 

\textbf{Data Transfer:} one of the primary challenges is the overhead associated with transferring data between the CPU and GPU. While the time required for copying data can take some time, especially when the bandwidth is not large enough to justify the transfer costs, which cannot be compensated then with the elapsed time of the GPU-accelerated workload. 

\textbf{Tile Size:} In the CPU version of \textit{sTiles}, the tile size is typically kept small, often determined by the L3 cache size, which helps optimize performance by reducing memory access times. On the GPU, however, much larger tile sizes can be used due to its ability to handle massive parallel workloads. Larger tiles allow for better use of the GPU’s resources by increasing hardware occupancy. For example, if the bandwidth of an arrowhead matrix is 3000 and the CPU tile size is set at 120, adding more CPU cores can improve performance scalability given the availability of the workloads. In contrast, GPU tile sizes are often larger (e.g., greater than 600) than CPUs', since
GPU tasks are inherently parallel while CPU tasks are sequential.

\textbf{GPU Memory:} Efficient memory management is critical when adapting \textit{sTiles} to GPUs, as the available memory on a GPU is limited compared to the CPU. Two main scenarios arise when dealing with matrix data on the GPU. In the first scenario, where the matrix fully fits into the GPU memory, computations can be performed directly on the matrix without the need for frequent data transfers between the CPU and GPU. This is the ideal case, as it minimizes the overhead associated with data movement, allowing the GPU to focus solely on computations, which maximizes performance. The second scenario is referred to as ``out-of-core." In this case, the matrix is too large to fit into the GPU memory, requiring data to be fetched from the CPU to the GPU more frequently. This case introduces additional complexity, as efficient memory management becomes critical to ensure that data transfers do not become a bottleneck. One approach is to enforce data reuse, by applying as many tasks as possible on tiles already loaded onto the GPU memory, before releasing them to the CPU and making space for subsequent tiles and their corresponding tasks~\cite{ren2024accelerating}. We will address this scenario in a future work.

\textbf{Implementation:} In the GPU prototype implementation of \textit{sTiles}, each core is assigned its own CUDA stream, allowing for concurrent execution of tasks across multiple streams. This ensures that different portions of the matrix can be processed simultaneously, maximizing the GPU's parallelism. The CPU kernels used for operations such as Cholesky factorization, matrix multiplication, and triangular solves are replaced by their GPU counterparts: \texttt{cusolverDnDpotrf} for Cholesky factorization, \texttt{cublasDsyrk} for symmetric rank-k updates, \texttt{cublasDtrsm} for triangular solves, and \texttt{cublasDgemm} for general matrix-matrix multiplications. 

The GPU version of \textit{sTiles} is particularly well-suited for arrowhead matrices that are not very sparse and have a large bandwidth, where the increased computational demand can benefit from the parallel processing power of the GPU. However, as the matrix size grows beyond the memory limits of a single GPU, future work could focus on extending the implementation to support multiple GPUs~\cite{Ltaief-vecpar} or addressing out-of-core problems~\cite{ren2024accelerating}. This would involve dynamically loading and offloading data between the CPU and GPU, allowing \textit{sTiles} to handle larger problem sizes more efficiently.

\section{Performance Comparison and Experimental Results} \label{sec4} 

In this section, we compare the performance of the \textit{sTiles} framework with other leading libraries for Cholesky factorization by conducting experiments on matrices of varying sizes, bandwidths, and arrowhead thicknesses to assess the scalability and efficiency of \textit{sTiles} in different scenarios. The tests are performed on two high-performance servers: Server 1, featuring an Intel(R) Xeon(R) Gold 6230R CPU (``Emerald Rapids'') running at 2.10GHz with 26 cores per socket, 2 sockets, and a 71.5MB L3 cache (2 instances), and Server 2, equipped with an AMD EPYC 7713 64-Core CPU (``Milan'') running at 1.5GHz with 64 cores per socket, 2 sockets, and a 512MB L3 cache (16 instances). These two servers, representing different hardware architectures (Intel and AMD), allow for an evaluation of the libraries' performance across diverse systems.

In our experiments, the impact of tile size on the performance of Cholesky factorization is evaluated across different architectures. Detailed results of this evaluation can be found in Appendix B: Tile Size Evaluation.

\begin{figure}%[htbp]
    \centering
    % First image
    \includegraphics[width=0.9\columnwidth]{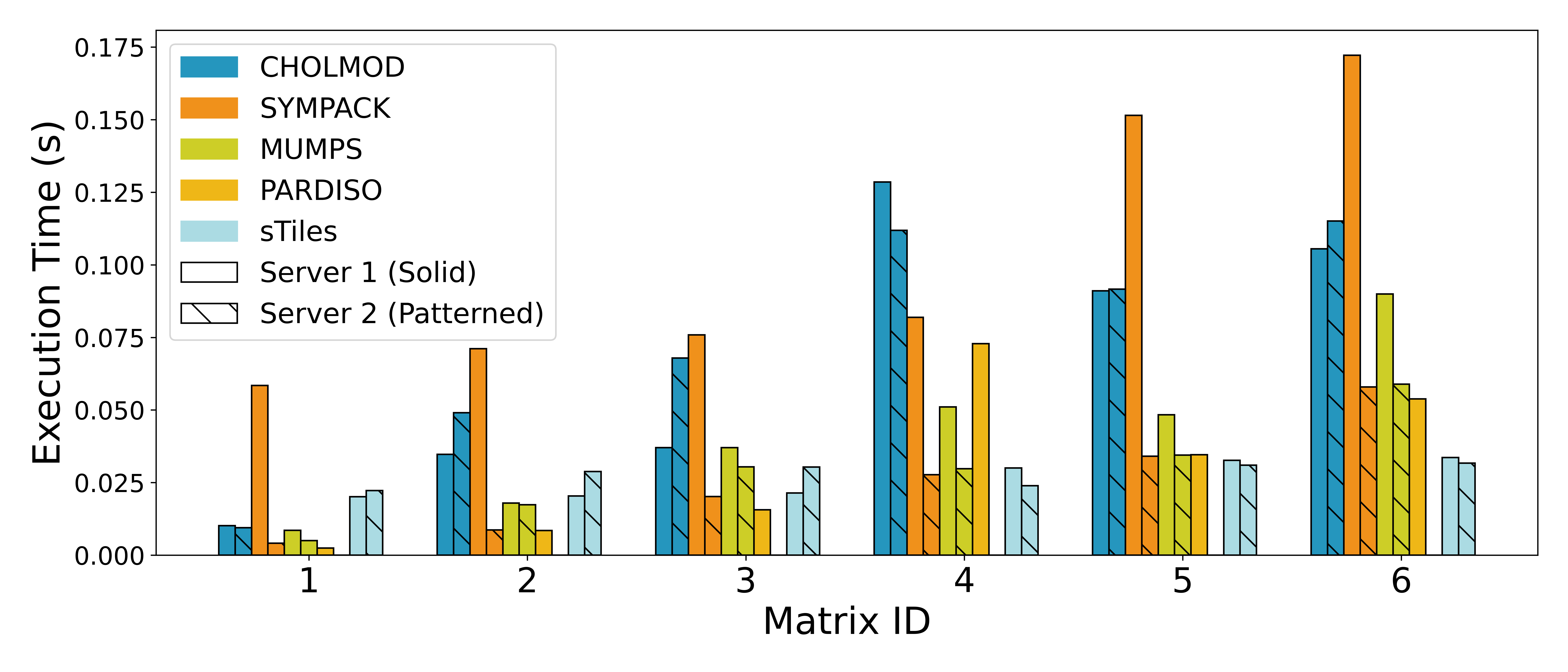} % Replace with your image
    %\caption{Caption for the first image}
    
    \vspace{0.001cm} % Optional: add vertical space between images
    
    % Second image
    \includegraphics[width=0.9\columnwidth]{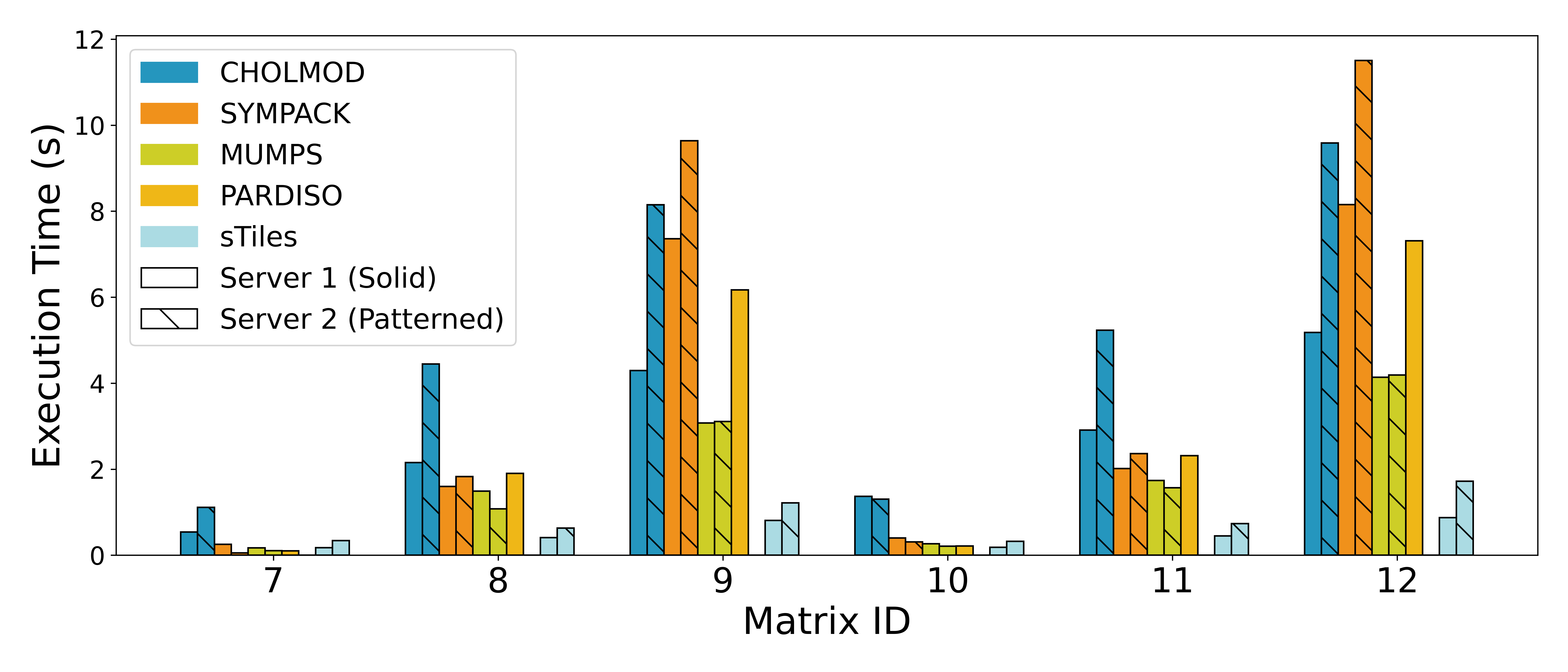} % Replace with your image
    %\caption{Caption for the second image}
    
    \vspace{0.001cm} % Optional: add vertical space between images
    
    % Third image
    \includegraphics[width=0.9\columnwidth]{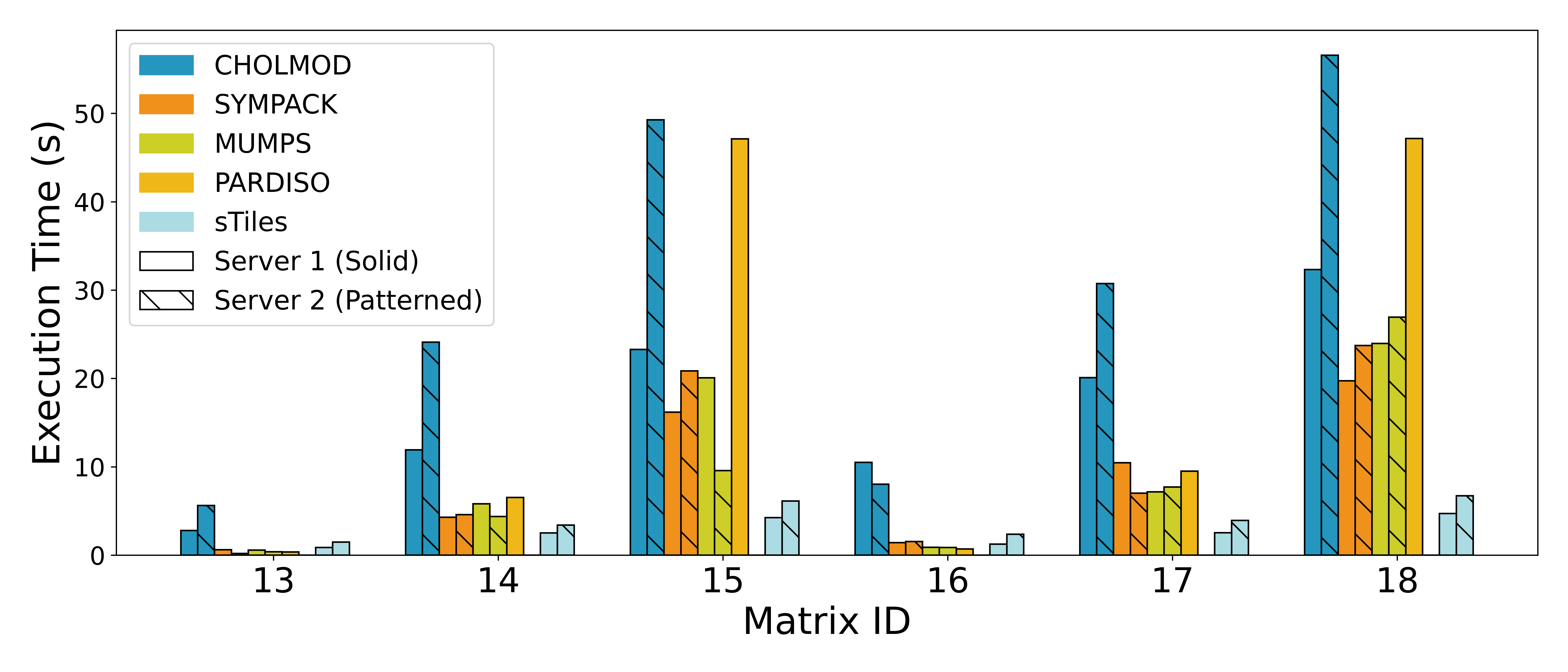} % Replace with your image
    \caption{Elapsed time of Cholesky factorization using different libraries on two servers.}
    \label{fig:firsexper}
\end{figure}

\subsection{Libraries for Comparison}

The performance of \textit{sTiles} is compared against several established libraries for Cholesky factorization:

\begin{enumerate}
    \item \textbf{CHOLMOD:} A free, shared-memory library for sparse symmetric positive-definite matrix factorization and linear system solving \cite{davis2005cholmod, 10.1145/1391989.1391995}. CHOLMOD utilizes LAPACK and BLAS for efficient factorization and automatically selects appropriate matrix reordering techniques such as AMD, CAMD, COLAMD, and CCOLAMD to minimize fill-in \cite{amestoy1996approximate}.
    
    \item \textbf{MUMPS:} A free, shared/distributed-memory solver designed for large sparse linear systems \cite{amestoy2000mumps}. We configured MUMPS to run in shared-memory mode with PETSc for handling sparse matrices. We evaluated multiple reordering strategies, including the AUTO and METIS options, with METIS providing better performance for arrowhead matrices.
    
    \item \textbf{SymPACK:} A free distributed-memory library optimized for high-performance sparse symmetric matrix factorization \cite{jacquelinsympack}. SymPACK implements a multifrontal parallel Cholesky factorization and we use METIS for reordering that gives better performance.
    
    \item \textbf{PARDISO:} A non-free solver for sparse linear systems, including symmetric and unsymmetric matrices \cite{schenk2001pardiso}. PARDISO employs both shared and distributed-memory parallelism and incorporates its own version of the METIS ordering algorithm to enhance reordering efficiency. We utilized a licensed version on server 1 for performance evaluation.
\end{enumerate}

\subsection{Matrix Properties}

The matrices used for the experiments are symmetric and positive definite, with varying sizes, bandwidths, arrowhead thicknesses, and sparsity levels. Notably, when the bandwidth is 100 or 1000, the diagonal part of the arrowhead matrix exhibits a block diagonal structure, meaning there is no correlation between the blocks. This structure significantly influences the behavior of the Cholesky factorization, as the absence of correlation between blocks reduces computational complexity in these regions. 

\begin{table}[!t] % Place the table at the top of the column
\centering
\footnotesize
\begin{tabularx}{\columnwidth}{c|X|X|X|X}
\textbf{ID} & \textbf{Size} & \textbf{Bandwidth} & \textbf{Arrowhead Thickness} & \textbf{Density (\%)} \\
\hline
\vspace{0.2ex} \\ % Add space above
\multicolumn{5}{c}{\textbf{Matrices for CPU Experiments}} \\[0.5ex] % Add vertical spacing around header
\hline
1  & 10,010   & 100   & 10  & 0.4083  \\
2  & 10,010   & 200   & 10  & 0.6051  \\
3  & 10,010   & 300   & 10  & 0.6434  \\
4  & 10,200   & 100   & 200 & 3.9380  \\
5  & 10,200   & 200   & 200 & 4.0325  \\
6  & 10,200   & 300   & 200 & 4.0666  \\
7  & 100,010  & 1000  & 10  & 0.1211  \\
8  & 100,010  & 2000  & 10  & 0.2199  \\
9  & 100,010  & 3000  & 10  & 0.2589  \\
10 & 100,200  & 1000  & 200 & 0.4988  \\
11 & 100,200  & 2000  & 200 & 0.5977  \\
12 & 100,200  & 3000  & 200 & 0.6370  \\
13 & 500,010  & 1000  & 10  & 0.0242  \\
14 & 500,010  & 2000  & 10  & 0.0441  \\
15 & 500,010  & 3000  & 10  & 0.0520  \\
16 & 500,200  & 1000  & 200 & 0.1001  \\
17 & 500,200  & 2000  & 200 & 0.1200  \\
18 & 500,200  & 3000  & 200 & 0.1281  \\
\vspace{0.5ex} \\ % Add space above
\multicolumn{5}{c}{\textbf{Matrices for CPU and GPU Experiments}} \\[0.2ex] % Add vertical spacing around header
\hline
19 & 50,010   & 15,000 & 10  & 0.3123  \\
20 & 1,000,010 & 3,000  & 10  & 0.0524  \\
\end{tabularx}
\caption{\footnotesize{Matrix properties used in Cholesky factorization experiments related to INLA framework \cite{van2023new}.}}
\label{tab:matrix_propertiessec5}
\end{table}

These matrices are generated within the context of statistical modeling and can arise from Kronecker products of a inverse covariance matrix representing temporal and spatial components. In many real-world applications, the arrowhead region in arrowhead matrices typically does not exceed 200 in thickness. For example, in INLA, the 200 columns often correspond to 200 fixed effects that need to be estimated, which is considered already a substantial number. By selecting arrowhead thicknesses of 10 and 200, we aim to capture the range of typical and more complex scenarios in such applications. The matrix properties are summarized in Table~\ref{tab:matrix_propertiessec5}.

\subsection{Cross-Architecture Performance Comparison}

We performed the Cholesky factorization on the matrices in Table \ref{tab:matrix_propertiessec5} using varying core configurations on both servers. On server 1 (Intel Xeon 6230R), the core counts were: 1, 2, 4, 8, 16, 32 and 52. On server 2 (AMD EPYC 7713), the core counts used were: 1, 2, 4, 8, 16, 32, 64, and 128. For each configuration, we measured the execution time and recorded the minimum time required to complete the Cholesky factorization (with tile size equals 120 for \textit{sTiles}), excluding the preprocessing phase. The number of cores used to achieve the best performance could vary between libraries. In this analysis, we plot the minimum execution time for each library at its optimal core count. The results of these experiments are presented Figure~\ref{fig:firsexper}.

For smaller matrices, particularly those with sizes around 10k, the execution time is already minimal across all libraries. In most cases, the performance differences between libraries are less significant. However, PARDISO and SymPACK shows strong performance for these smaller matrices. In cases where the diagonal part of the arrowhead matrix is composed of independent block diagonals (Matrices 1, 7, 10, 13, 16), some libraries outperformed \textit{sTiles}. This is primarily due to the structure of the matrix. \textit{sTiles} can be further optimized in these cases if the tile size is chosen to match the size of the small blocks in the diagonal part or is proportional to it. For larger matrices, especially those with sizes around 100k and 500k, the performance of the libraries observed on server 1 (Intel Xeon) is generally superior to that on server 2 (AMD EPYC). MUMPS shows strong performance for matrices around 100k, while SymPACK performs well for matrices around 500k. Overall, \textit{sTiles} is better than other libraries in most cases. This is particularly evident in cases where the bandwidth becomes higher with a thicker arrowhead region, such as in matrices 6, 9, 12, 15, and 18.

\subsection{Performance Scalability Analysis}

\begin{figure}%[htbp]
    \centering
    \includegraphics[width=\columnwidth]{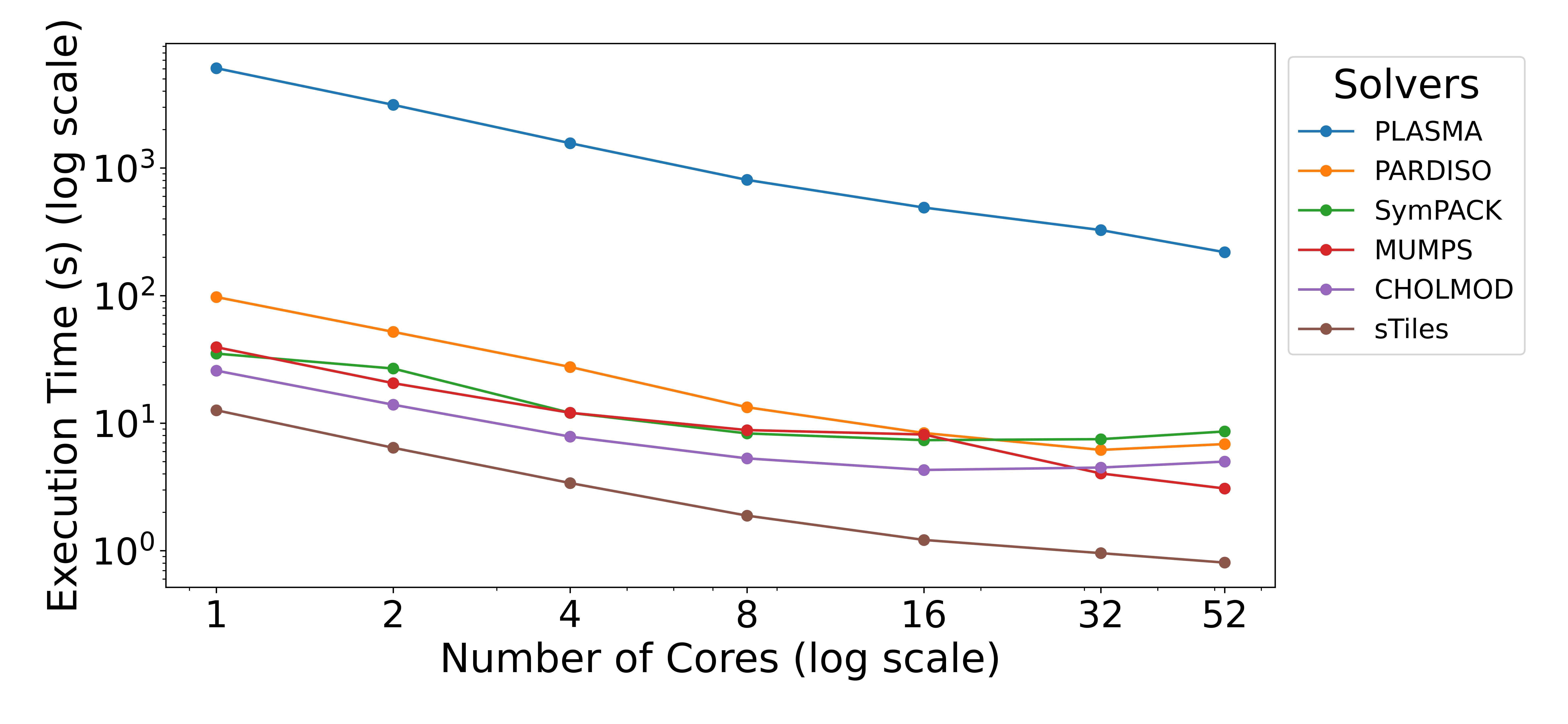}
    \caption{Performance scalability for Matrix ID 9 using different libraries.}
    \label{coresvar}
\end{figure}

The scalability of the libraries was evaluated using Matrix ID 9 on server 1 (Intel Xeon 6230R) across various core counts (1, 2, 4, 8, 16, 32 and 52), as shown in Figure~\ref{coresvar}. \textit{sTiles} exhibited the best scalability, with consistent reductions in execution time as the core count increased, particularly up to 32 cores, where it significantly outperformed other libraries. Beyond 32 cores, the performance gains diminished but remained superior, demonstrating its efficient handling of computationally intensive tasks. CHOLMOD and PARDISO scaled reasonably well at lower core counts but saw diminishing returns beyond 16 and 32 cores, respectively, likely due to memory bottlenecks and limited parallelization strategies. SymPACK performed well up to 32 cores but struggled to maintain efficiency at higher core counts, while MUMPS showed strong initial performance but also plateaued as the core count increased. To provide broader context we added PLASMA in the comparison. While it exhibited strong scalability, its execution times were significantly higher than the sparse-specific libraries due to the overhead of dense computations, highlighting the trade-off between general-purpose and specialized approaches like \textit{sTiles}. Overall, \textit{sTiles} proved the most scalable, particularly for larger core configurations.

The scalability trends observed for Matrix ID 9 were similar across other matrices, and the best achieved time for each library using different core configurations has already been presented in the previous section.

\subsection{Tree Reduction Performance in Cholesky Factorization}

\begin{figure}%[htbp]
    \centering
    \includegraphics[width=0.7\columnwidth]{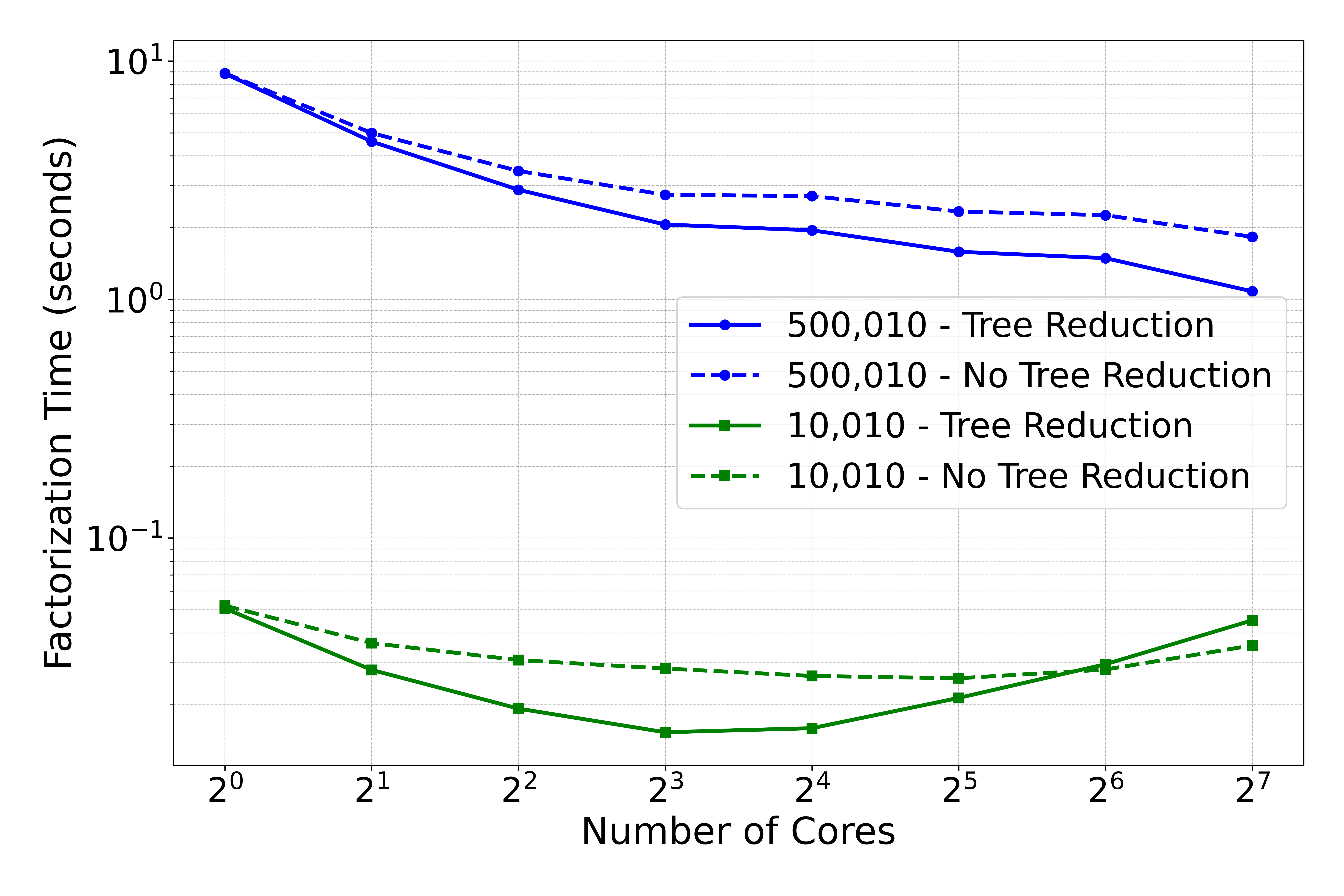}
    \caption{Performance of tree reduction in Cholesky factorization for matrices of different sizes.}
    \label{fig:tree_reduction_results}
\end{figure}

To evaluate the performance of tree reduction in Cholesky factorization, we selected matrix IDs 2 and 14 as test cases. These matrices were chosen to represent varying levels of SYRK or GEMM accumulation, which is influenced by both matrix size and the presence of an arrowhead region in the arrowhead matrices. With a tile size of 120, matrix ID 2 exhibits 84 accumulations, while matrix ID 14 shows 4,166 accumulations. The experiments were conducted using 2, 4, 8, 16, 32, 64, and 128 cores on Server 2 to measure the Cholesky factorization time. The results, comparing the performance with and without tree reduction, are presented in Figure \ref{fig:tree_reduction_results}.

The analysis of factorization performance across matrices of sizes 500,010; and 10,010 reveals distinct patterns. For the 500,010 matrix, tree reduction significantly improves performance, especially as core counts increase, showing the best scalability. As the core count rises, execution time with tree reduction continues to drop, reaching 1 second at 128 cores compared to 1.8 seconds without it. For the 10,010 matrix; tree reduction still provides performance gains, but diminishing returns become apparent beyond 16 cores as communication overhead increases. This effect is particularly pronounced because the number of cores exceeds the number of SYRK or GEMM accumulations, which is 84. At 64 and 128 cores, the computational workload becomes insufficient to keep all cores fully utilized, leading to performance stabilization or slight increases due to synchronization overhead.

\textit{Our recommendation is to use tree reduction when there are at least 2 cores, and when the number of accumulations is at least double the number of cores being used, particularly in larger matrices where there is sufficient computational workload.}

\subsection{GPU Acceleration Example}

To explore the GPU capabilities of the STiles framework, we evaluated two large matrices (IDs 19 and 20) from Table \ref{tab:matrix_propertiessec5}. These matrices were chosen to test performance under extreme conditions and assess the impact of large bandwidths on factorization.

For GPU execution, we set the tile size to 600, which we determined as the optimal choice after a series of experiments. For comparison, the CPU version of \textit{sTiles} was configured with a tile size of 120, following the default configuration used in previous experiments. The experiments were conducted using a single NVIDIA A100-SXM4 GPU (1.16 GHz, 80 GB HBM2 memory, ``Ampere'') and 32 AMD EPYC 7713 CPU cores (1.99 GHz, 512 GB system RAM, ``Milan'').

The results are summarized in Table \ref{tab:side_by_side_results}, which shows the execution times (in seconds) for both the CPU and GPU implementations across varying core counts. These results clearly demonstrate the advantage of GPU acceleration in \textit{sTiles}. For the smaller 50,010-sized matrix, the GPU implementation achieves a dramatic reduction in execution time, with over a 5X speedup compared to the 32-core AMD CPU. This speedup is within the range of what the roofline performance model says for both CPU and GPU systems (1Tflop/s vs 20Tflops/s) from a theoretical peak performance perspective. However, the host-device data movement hinders the speedup and the absolute performance, which does not get compensated by the workload given the relatively limited concurrency. 
Similarly, for the larger 1,000,010-sized matrix, the GPU demonstrates a 2.7X speedup compared to the CPU on 32 cores, a lower speedup than for the small matrix case though, due to the widening gap of the data motion/compute (or surface/volume) ratio. While the GPU hardware does bring performance benefits, the performance gains also stem from the preprocessing during factorization, where differences in tile size result in varying tiled matrix structures, symbolic factorizations, and task distributions.
 
\begin{table}%[h!]
\centering
\small
\renewcommand{\arraystretch}{1.02} % Adjust row spacing
\resizebox{\columnwidth}{!}{ % Adjust the size to fit the page
\begin{tabular}{c|c|c|c|c}
\multicolumn{1}{c|}{\textbf{Cores/Streams}} & \multicolumn{2}{c|}{\textbf{Matrix Size: 50,010}} & \multicolumn{2}{c}{\textbf{Matrix Size: 1,000,010}} \\[1.5ex]
 & \textbf{Time (CPU)} & \textbf{Time (GPU)} & \textbf{Time (CPU)} & \textbf{Time (GPU)} \\[1.5ex]
\hline
1   & 97.66  & 1.46  & 239.73 & 9.24  \\
2   & 48.96  & 1.09  & 106.18 & 6.61 \\
4   & 24.91  & 0.94  & 60.18  & 5.89 \\
8   & 12.68  & 0.86  & 31.05  & 5.75 \\
16  & 6.53   & 0.79  & 20.12  & 5.76 \\
32  & 4.07   & 0.81  & 15.95  & 5.75 \\
\end{tabular}
} % End of resizebox
\caption{Execution times for Matrix ID 19 and Matrix ID 20, including GPU data movement and computation times.}
\label{tab:side_by_side_results}
\end{table}

The performance gap between CPU and GPU is more pronounced in scenarios where the matrix size and bandwidth are larger, as the GPU's parallel processing and memory bandwidth become more effective in handling the computational load. Interestingly, the GPU performance remains relatively stable across core configurations, suggesting that the GPU efficiently handles the workload with minimal scaling dependency. Overall, these experiments validate the effectiveness of GPU acceleration in \textit{sTiles}, particularly for large, bandwidth-intensive matrices. Future work could explore multi-GPU setups and optimizations for specific matrix structures to further enhance performance.

\subsection{Balanced Efficiency for Sparse and Dense Approaches}

\begin{figure}%[htbp]
    \centering
    \includegraphics[width=0.7\columnwidth]{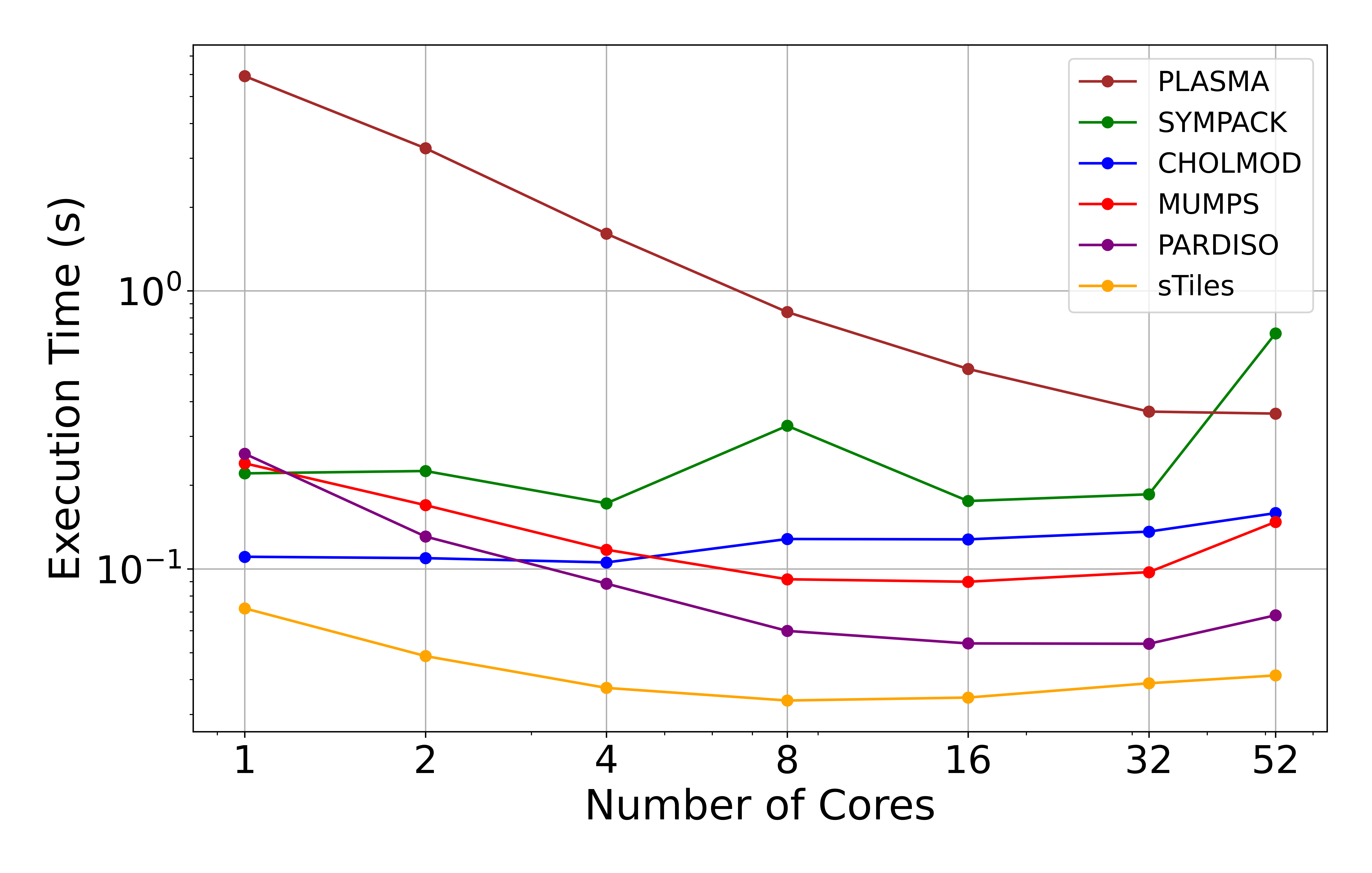}
    \caption{Execution times for Matrix ID 6 using different libraries across varying core counts.}
    \label{alllibs}
\end{figure}

We selected Matrix ID 6, a relatively small matrix, as larger matrices would not be feasible for comparison with PLASMA, given its fully dense factorization approach. In this section, we present the execution times for Matrix ID 6 across varying core counts, extending our earlier analysis by including PLASMA. While PLASMA processes all data as dense, \textit{sTiles} offers a balanced approach, efficiently managing both sparse and dense regions within the matrix. This comparison, illustrated in Figure~\ref{alllibs}, highlights \textit{sTiles}’ flexibility and competitive performance alongside sparse-oriented libraries such as CHOLMOD, SymPACK, MUMPS, and PARDISO, demonstrating its ability to perform well across different core configurations.

Additionally, scalability analysis is not the primary focus in this case, as PLASMA, which treats the entire matrix as dense, does not fully capitalize on the sparsity patterns present in the arrowhead matrices. These matrices typically have a critical path, limiting the extent of parallelism achievable in such structures.

\section{Conclusion}
\label{sec5}
In this paper, we introduce \textit{sTiles}, a hybrid Cholesky factorization framework that balances computational efficiency for arrowhead matrices. Traditional sparse methods excel in handling highly sparse matrices but struggle with dense regions, while fully dense methods are inefficient for matrices with large sparse areas. \textit{sTiles} fills this gap by efficiently targeting matrices that are neither purely sparse nor fully dense, making it applicable to a broader range of scientific and engineering problems.

We presented a comprehensive framework, beginning with preprocessing analysis that optimizes matrix structures through permutation techniques and proceeds to parallel factorization strategies. Our experimental results demonstrate that \textit{sTiles} outperforms other leading libraries, such as \textit{CHOLMOD}, \textit{SymPACK}, \textit{MUMPS}, and \textit{PARDISO}, across a range of matrix sizes and computational environments. Additionally, we explored the potential of GPU acceleration, where significant performance gains were observed, especially for matrices with large bandwidths.

Future work will focus on extending the framework to support multi-GPU configurations, further optimizing tile sizes for specific matrix structures, and exploring dynamic memory management strategies for out-of-core computations on GPUs \cite{ren2024accelerating}. \textit{sTiles} provides a robust foundation for efficient matrix factorization in large-scale Bayesian inference like INLA and other scientific computing fields, offering a balanced approach to handling sparse and dense matrix regions with substantial computational benefits.

\newpage
\section*{Appendix A: Multiple Concurrent Cholesky Factorizations}
\label{appendix_concurrent}

\textbf{Multi-level Execution:} Despite the implementation of parallel strategies in \textit{sTiles}, simply increasing the number of cores does not always yield substantial performance gains. This is primarily due to dependencies in matrix operations, for an arrowhead structure, which constrain the extent to which tasks can be parallelized. However, a notable opportunity for parallelism arises when performing multiple independent Cholesky factorizations concurrently.

For instance, when computing gradients using the central difference method for a convex function \( f(\pmb{\theta}) \), where the parameter vector \( \pmb{\theta} \) has dimension \( n \), each gradient component requires two function evaluations: \( f(\theta_i + h) \) and \( f(\theta_i - h) \) for each \( i \), with \( h \) as the step size. Since these function evaluations, each involving a Cholesky factorization, are independent, the process can be parallelized. This allows for \( 2n \) Cholesky factorizations to be executed in parallel, leading to a significant improvement in resource utilization and computational efficiency. One of the key applications of concurrent Cholesky factorizations is the INLA method, where factorizations are needed at various stages of the Bayesian inference process. Examples include the computation of the Smart Gradient \cite{fattah2022smart} and Parallel Line Search techniques \cite{gaedke2023parallelized}, both of which benefit from concurrent execution to enhance performance and scalability.

\begin{figure}[htbp]
    \centering
    \subfloat[NUMA node core IDs layout.]{\includegraphics[width=0.45\columnwidth]{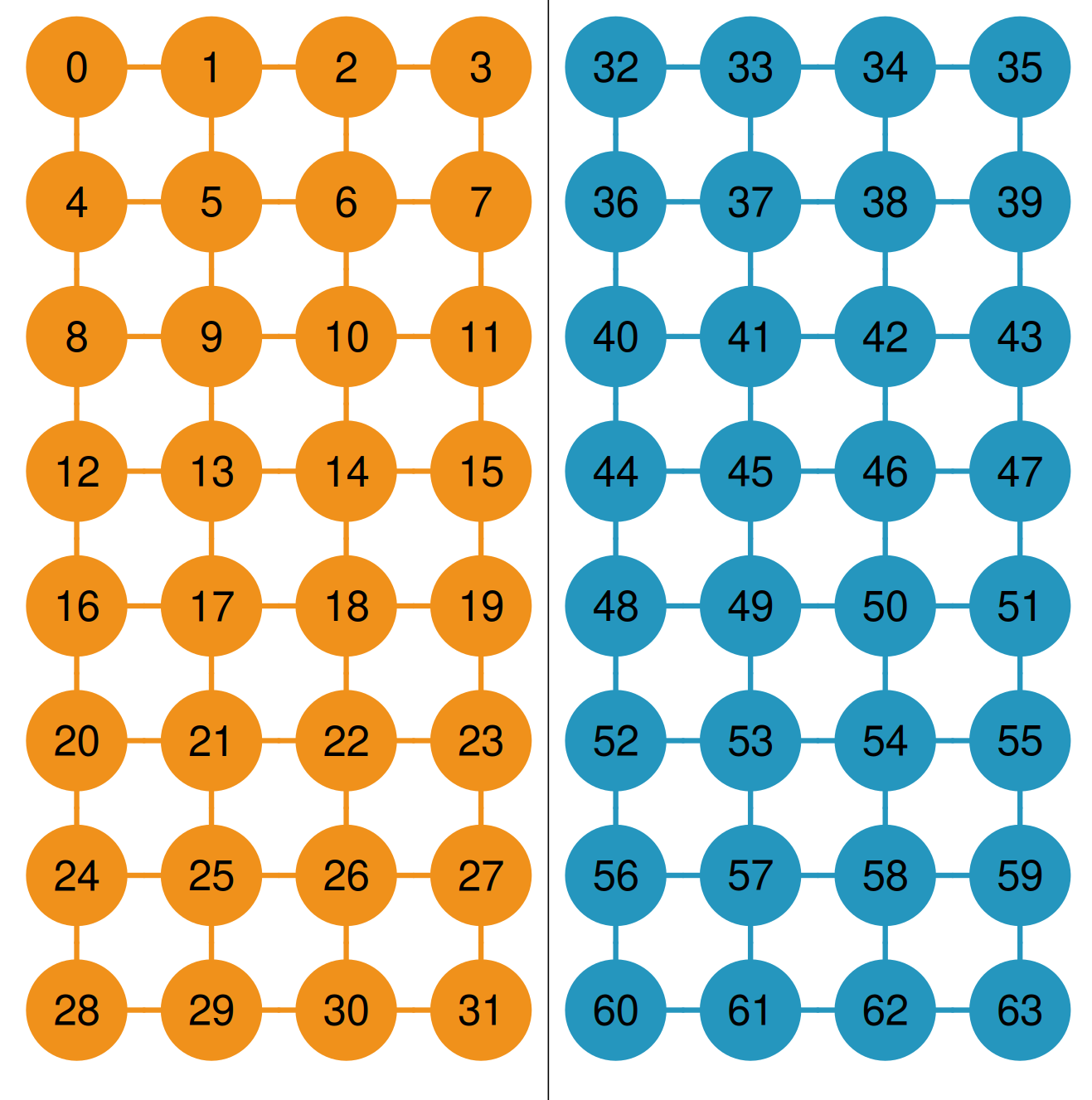}\label{fig:numa_layout1}}\hspace{0.05\columnwidth}
    \subfloat[Core assignment for four Cholesky factorizations.]{\includegraphics[width=0.45\columnwidth]{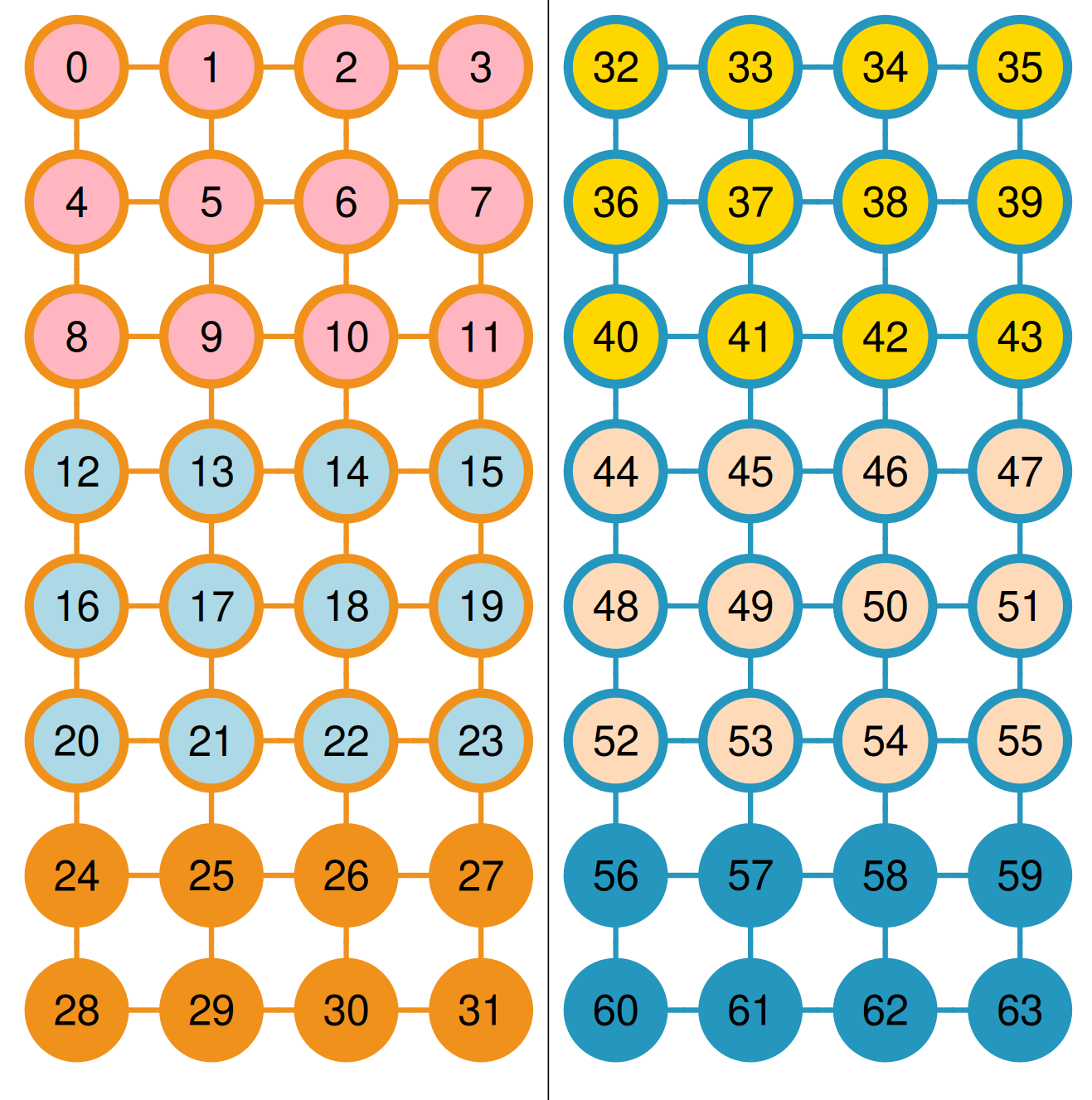}\label{fig:numa_layout2}}
    \caption{NUMA layout and core assignment for Cholesky factorizations.}
    \label{fig:numa_layouts}
\end{figure}

This parallelism can be achieved using shared-memory (OpenMP) or distributed-memory (MPI) parallelism. Shared memory is particularly useful when each Cholesky factorization does not require many cores, allowing for efficient parallel execution within a single node. On the other hand, when the Cholesky factorization involves a substantial workload, the current implementation can fully utilize the cores on one node and, if necessary, run additional factorizations on separate nodes. This setup ensures efficient scaling across multiple nodes. Moreover, the current approach could be extended to allow a single Cholesky factorization to be distributed and computed across multiple nodes using nested dissection ordering, further enhancing performance for larger matrices. 

\textbf{NUMA-Aware Core Binding:} When running parallel Cholesky factorizations on a machine with more than one NUMA node, it is important to bind the cores according to their NUMA node CPU IDs. Binding the cores for a Cholesky factorization to a single NUMA node can reduce latency and improve memory access efficiency. In machines where the NUMA nodes have more complex configurations, as shown in Figure~\ref{fig:numa_layout1} (NUMA node 0 CPU(s): 0-31 and NUMA node 1 CPU(s): 32-63), we carefully bind the cores so that they remain on the same NUMA node for each Cholesky factorization. For example, in Figure~\ref{fig:numa_layout2}, four independent Cholesky factorizations are distributed across two NUMA nodes, with each factorization using 12 cores. The orange and blue colors represent the cores allocated from NUMA node 0 and NUMA node 1, respectively, while the other colors represent the core IDs assigned to each Cholesky call.
While the PLASMA library automatically binds cores for a single Cholesky factorization, assuming it can take all available cores on a node, our implementation in \textit{sTiles} differs. In \textit{sTiles}, we run multiple independent Cholesky factorizations concurrently, so we must manually bind cores in a way that ensures each Cholesky factorization uses the resources of a single NUMA node whenever possible. This allows for better memory locality and more efficient execution across multiple Cholesky factorizations.

\section*{Appendix B: Tile Size Evaluation}
\label{appendix_tile_size_evaulation}

In this experiment, we evaluate the impact of varying tile sizes on the performance of Cholesky factorization across different architectures using a matrix of size 100,200 (Matrix ID 12). Specifically, we assess how tile size influences execution time and cache efficiency. The experiments were conducted using 4 cores on the servers 1 and 2, with tile sizes ranging from 40 to 400.

Matrix ID 12 was chosen for this evaluation because it is large enough to represent a substantial computational workload, allowing for a meaningful assessment of performance variations across different tile sizes. Additionally, it features a thick arrowhead structure with a 200-column width, providing a more complex scenario that tests the efficiency of Cholesky factorization under increased computational demands. Larger matrices exhibit similar behavior, making Matrix ID 12 a representative choice for this analysis.

\begin{figure}[hbt!]
    \centering
    \begin{subfigure}[b]{\columnwidth}
        \centering
        \includegraphics[width=\columnwidth]{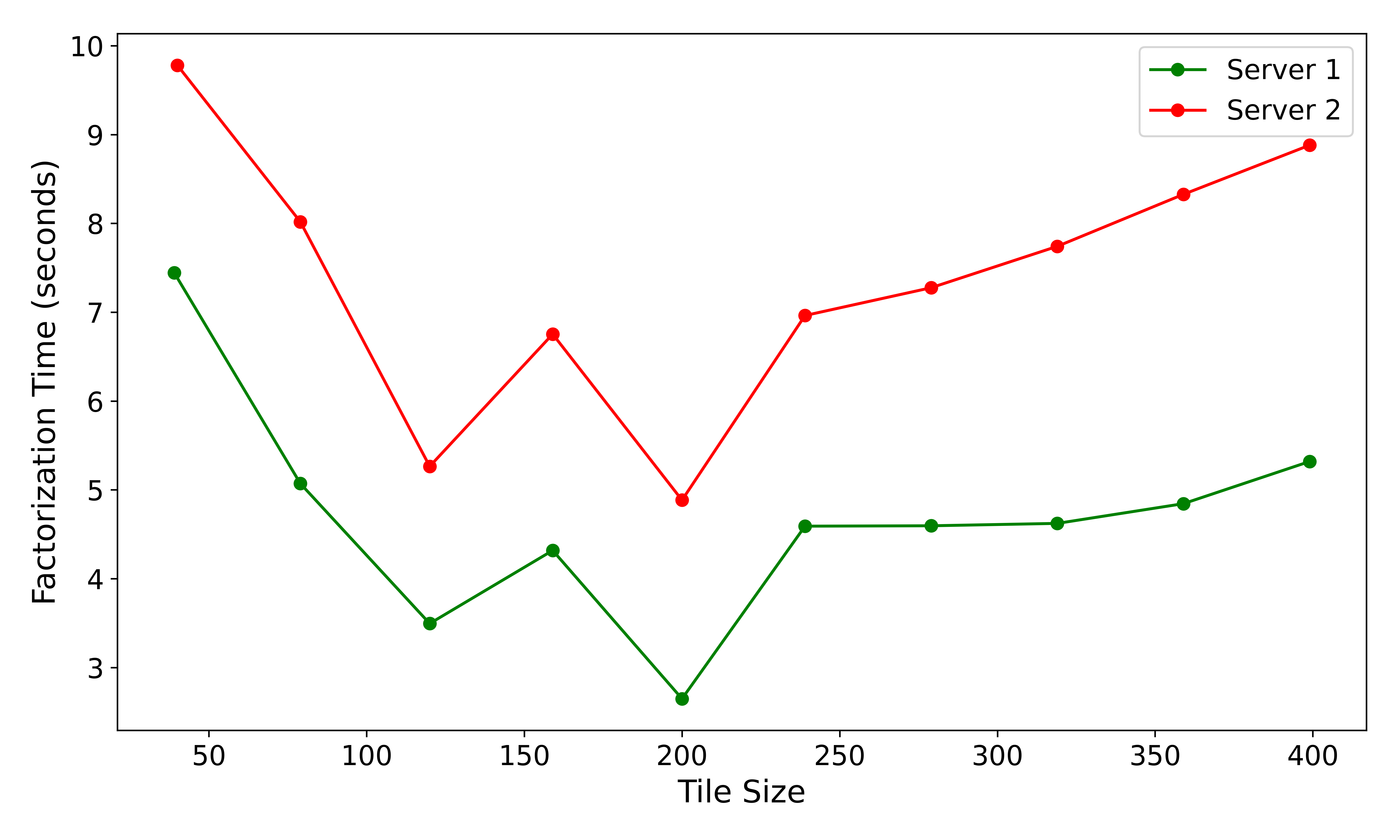}
    \end{subfigure}
    \\[0.05cm]
    %\rule{0.5\columnwidth}{0.1pt}
    %\\[0.1cm]
    \begin{subfigure}[b]{\columnwidth}
        \centering
        \includegraphics[width=\columnwidth]{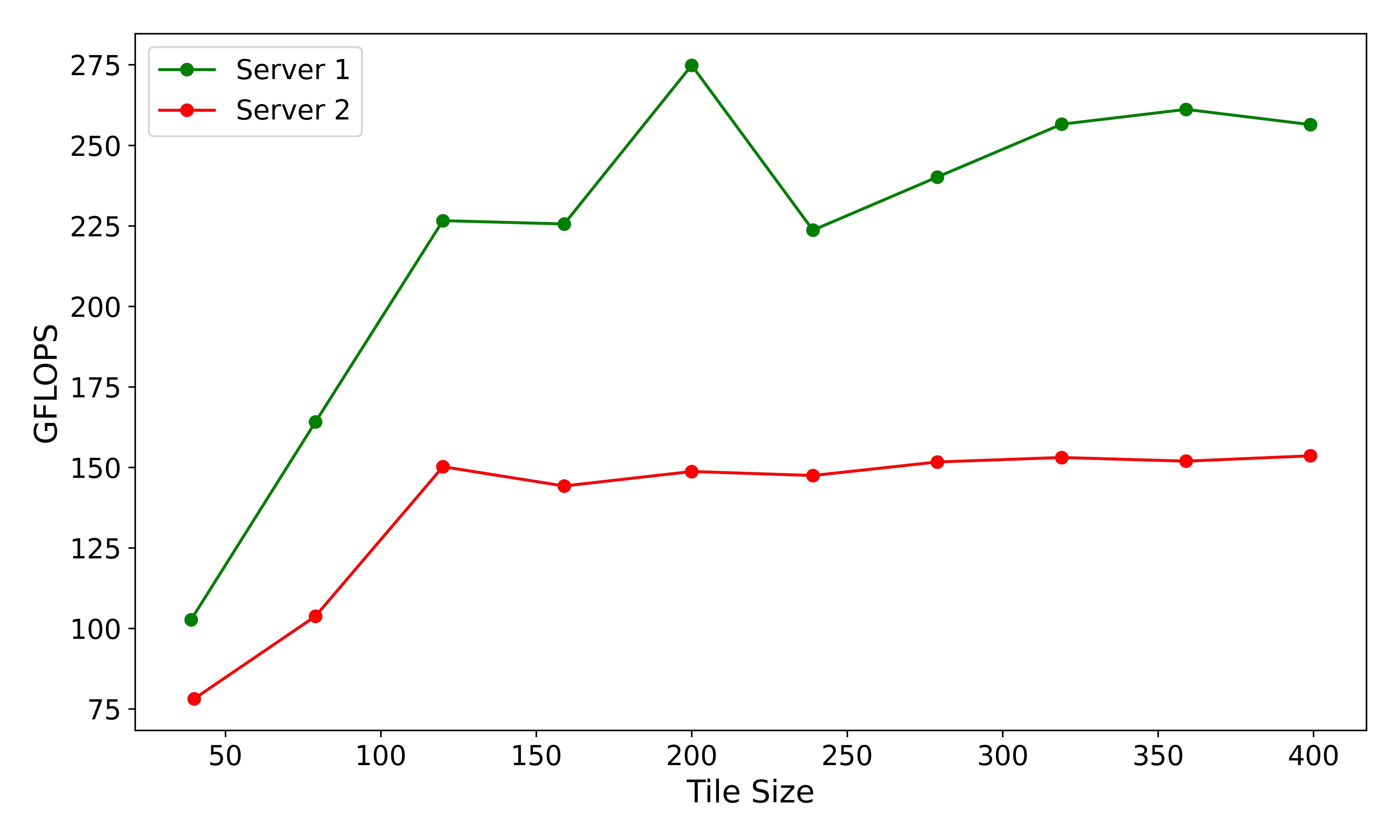}
    \end{subfigure}
    \caption{Execution times for Matrix ID 9 using different libraries across varying core counts.}
    \label{fig:tilesizeexp}
\end{figure}

Previous studies, particularly with PLASMA, have extensively investigated tile size tuning for dense matrix factorizations. PLASMA identified that a tile size of 120 strikes a good balance between computational efficiency and memory access times for Cholesky factorization. Our experiments, however, focus on determining whether this optimal tile size holds when performing Cholesky factorization on arrowhead-structured matrices using \textit{sTiles}. The results for factorization time and GFLOPS (floating-point operations per second) for each tile size on the three systems are shown in Figure~\ref{fig:tilesizeexp}. To better understand the relationship between tile size and performance, we categorize the results into three key ranges: smaller tile sizes, optimal tile sizes, and larger tile sizes. 

\begin{itemize}
    \item \textbf{Smaller Tile Sizes (40-80):} These smaller tile sizes introduce overhead due to frequent memory transfers, especially in sparse matrix scenarios where memory access locality is harder to maintain. As a result, execution times increase across all systems.

    \item \textbf{Optimal Tile Size (120-240):} As anticipated, tile sizes in the range of 120 to 240 yield the best performance. A tile size of 200 performs particularly well in cases where the matrix has a bandwidth is around 200 or when the number of columns in the arrowhead region aligns with this size. This alignment allows for more efficient processing of the dense blocks in the arrowhead region, leading to improved performance.

    \item \textbf{Larger Tile Sizes (240-400):} Increasing the tile size beyond 240 leads to performance degradation. Larger tile sizes result in inefficient cache utilization and increased memory access times, particularly on Server 2, which has a smaller L3 cache per core compared to other systems.
\end{itemize}

The wiggly pattern observed in Figure~\ref{fig:tilesizeexp} across different tile sizes could be attributed to the way elements are mapped to tiles during the computation. Some tile sizes may align more conveniently with the structure of the matrix, resulting in better cache locality and more efficient memory access, while others might lead to less optimal mappings, causing increased memory transfers and slower performance. To handle this variability, we keep the default tile size as recommended by PLASMA, which is set to 120. However, users are encouraged to experiment with different tile sizes and adjust this value according to the specific structure and characteristics of their matrices. 

\end{document}